\documentclass[12pt, draftclsnofoot, onecolumn]{IEEEtran}
\usepackage{indentfirst}
\usepackage{multirow}
\usepackage{graphicx}
\usepackage{color}
\usepackage{xcolor}
\usepackage{colortbl}
\definecolor{mypink}{rgb}{.99,.91,.95}
\usepackage{listings}
\usepackage{enumerate}
\usepackage{subfigure}
\usepackage{amsmath,amsthm,amscd,amssymb,latexsym,upref,stmaryrd}
\usepackage[noend]{algpseudocode}
\usepackage{algorithmicx,algorithm}
\usepackage{multirow} 
\usepackage{xcolor}
\usepackage{amsfonts,hyperref,epsfig}
\usepackage{CJK}
\usepackage{float}
\usepackage{cite}
\usepackage{subfigure}
\usepackage{lipsum}
\usepackage{multicol}
\usepackage{setspace}
\usepackage{bm}
\usepackage{stfloats}
\usepackage{booktabs}
\usepackage{threeparttable}

\newtheorem{Def}{Definition}%

\def\BibTeX{{\rm B\kern-.05em{\sc i\kern-.025em b}\kern-.08em
		T\kern-.1667em\lower.7ex\hbox{E}\kern-.125emX}}

\newtheorem{remark}{Remark}

\begin{document}
	\title{Enabling Plug-and-Play and Crowdsourcing SLAM in Wireless Communication Systems}

	\author{Jie~Yang, Chao-Kai~Wen, Shi~Jin, and Xiao~Li
		\thanks{Jie~Yang, Shi~Jin, and Xiao~Li are with the National Mobile Communications Research Laboratory, Southeast University, Nanjing, China (e-mail: \{yangjie;jinshi;li\_xiao\}@seu.edu.cn). Chao-Kai~Wen is with the Institute of Communications Engineering, National Sun Yat-sen University, Kaohsiung, 804, Taiwan (e-mail: chaokai.wen@mail.nsysu.edu.tw). 
		}
	}
	
	\maketitle	
	
	\begin{abstract}
		
		Simultaneous localization and mapping (SLAM) during communication is emerging. This technology promises to provide information on propagation environments and transceivers' location, thus creating several new services and applications for the Internet of Things and environment-aware communication. Using crowdsourcing data collected by multiple agents appears to be much potential for enhancing SLAM performance. However, the measurement uncertainties in practice and biased estimations from multiple agents may result in serious errors. This study develops a robust SLAM method with measurement plug-and-play and crowdsourcing mechanisms to address the above problems. First, we divide measurements into different categories according to their unknown biases and realize a measurement plug-and-play mechanism by extending the classic belief propagation (BP)-based SLAM method. The proposed mechanism can obtain the time-varying agent location, radio features, and corresponding measurement biases (such as clock bias, orientation bias, and received signal strength model parameters), with high accuracy and robustness in challenging scenarios without any prior information on anchors and agents. Next, we establish a probabilistic crowdsourcing-based SLAM mechanism, in which multiple agents cooperate to construct and refine the radio map in a decentralized manner. Our study presents the first BP-based crowdsourcing that resolves the ``double count" and ``data reliability" problems through the flexible application of probabilistic data association methods. Numerical results reveal that the crowdsourcing mechanism can further improve the accuracy of the mapping result, which, in turn, ensures the decimeter-level localization accuracy of each agent in a challenging propagation environment.
		
	\end{abstract}


	\section{Introduction} \label{sec:introduction}
	\IEEEPARstart{F}{uture} mobile communication systems will evolve into a brilliant and software-reconfigurable functionality paradigm that can provide ubiquitous communication and the perception, control, and optimization  of wireless environments \cite{6G1}. 
	Toward this end, simultaneous localization and mapping (SLAM) during communication has become an increasingly critical function that enables the applications of robotics, autonomous driving, and indoor localization.
	Ascertaining the locations of agents (or user equipment) and obtaining a map of the propagation environment can also significantly improve the communication rate and reliability of future wireless communication systems \cite{BP2,Location-aware}.
	However, obtaining accurate and reliable information about agents' locations and their propagation environment in wireless communication systems faces the following challenges, especially in scenarios with severe multipath effect:
	(a) \emph{practicability requirements},
	for which imprecise synchronization, an unknown heading direction, and other unknown measurement biases caused by hardware impairments may generate serious errors in localization;
	(b) \emph{data association ambiguities}, for which corrupted or missing measurements caused by complicated multipath propagation environments and the scalability of the number of features in the radio map may lead to data association ambiguities; and
	(c) \emph{multi-agent cooperation}, for which biased estimations from multiple agents may result in data inconsistency, thereby preventing crowdsourcing data from gaining benefits from cooperatives.

	Radio frequency (RF) signals have been widely used for locating mobile devices.
	In particular, \cite{UWB} is conducted according to the range measurements extracted from ultra-wideband signals. Wireless fidelity (Wi-Fi) is popularly
	applied to provide signals of opportunity for localization \cite{SOO} because of its easy and cost-effective deployment. Mobile devices can listen to beacon frames broadcast from nearby Wi-Fi access points, thereby obtaining received signal strength (RSS).
	RSS has been employed as a primary source for positioning solutions in line with trilateration or fingerprinting methods \cite{RSS}. 
	However, RSS is affected by many factors, such as small-scale fading and body shadowing, resulting in the degradation of localization quality.
	For this reason, many studies have focused on fine-grained channel state information (CSI) \cite{beaconCSI}.
	Distinguishable multipath components (MPCs) with high resolution of the angle of arrival (AOA), angle of departure (AOD), time of arrival (TOA), and frequency of arrival (FOA) can be extracted from the CSI, especially in high-frequency bands with large antenna arrays and highly directional transmission, such as millimeter-wave (mmWave) multiple-input multiple-output (MIMO) communication systems.
	Then, highly-accurate localization can be achieved without installing additional dedicated infrastructure \cite{loc1,mmwave1,mmwave2,loc2,slam2}.
	As mmWave signals follow quasi-optical propagation patterns \cite{mmwave}, MPCs arising from specular reflections are used in \cite{slam2,BP2}, and all the other diffuse MPCs are modeled as interference.

	Although localization using RF signals is promising, the resulting performances degrade in practice because of the measurement biases. 
	For example, each anchor (or base station) may use different transmission power levels, and the surrounding environment of each anchor may vary. Therefore, the RSS model parameters of each anchor are distant and unknown. 
	Moreover, the clock and orientation biases are also unknown, which deteriorate the localization performance on the basis of the TOA and AOA measurements, respectively.
	A feasible solution to solve the practical issues 
	entails using multiple shots-based methods (utilizing measurements obtained in successive time slots)
	and integrating multi-domain
	information collected from RF signals and sensors.

	SLAM takes advantage of the unchanging nature of the location and state of landmarks in the environment. 
	Through accumulating measurements in consecutive time slots, the landmark can be estimated by employing numerous measurements, thereby extending the single-shot solution to a multiple-shot one.
	Different branches of SLAM are overviewed in \cite{slam-survey}.
	The majority of classical SLAM approaches rely on a priori information of the environment.
	Loop closure is applied in SLAM algorithms to recognize a previously mapped place \cite{lc1,lc,8920098}.
	A low-cost solution is proposed in \cite{llm} for the automatic generation of a precise lane-level map concerning the local map segments.
	Recently, in-depth research has been conducted on communication-driven localization and mapping \cite{CSLAM, CSLAM2,CSLAM3,CSLAM4,slam2, BP-SLAM,BP2 }.
	The state and location of anchors are assumed known in \cite{CSLAM2,CSLAM3,CSLAM4,slam2}, or effectively-known in \cite{BP-SLAM,BP2}. 
	Robust localization from radio signals in indoor scenarios is challenging because of the complex multipath propagation and high probability of false alarm and missed detection.
	The feature-based SLAM algorithm proposed in \cite{BP-SLAM,BP2,slam2}, which performs sequential estimation of the states of a mobile agent and of ``potential features" characterizing the map, has low computational complexity and can cope with clutter, missed detections, and  data associations.
	However, the SLAM algorithm should not have the anchors' locations in advance to be viable for practical communication systems. The anchors' state may be flexible, e.g., the anchors may be temporarily connected or disconnected.
	Therefore, realizing SLAM in such challenging scenarios, in which anchors and agents' prior information is unavailable, is worth exploring.

	Most of the existing research on SLAM only considers a single agent.
	However, the connectivity of people and machines continues to increase in the age of the Internet of Everything, thereby providing the possibility for agents to cooperate in the process of SLAM.
	A centralized multi-agent collaborative mapping and
	positioning approach is proposed in \cite{mslam}.
	Under the crowdsourcing framework, location users have also become radio map providers, which can avoid labor-intensive data collection \cite{crowdsourcing,crowdsourcing-wifi}. 
	For centralized crowdsourcing approaches, communications are all directed to one entity, thereby agglomerating and fusing all data, and then sending the SLAM result back to the agents. 
	Recent trends move these processes from one entity to the cloud to take advantage of the available processing power. 
	Decentralized systems assume that each agent can build its decentralized map while communicating with the other agents or an entity.
	One of the main challenges is the ``double count" phenomenon arising from the inconsistency caused by the repeated counting of landmarks in the estimation process based on decentralized crowdsourcing.
	Another difficulty involves selecting the most valuable data and filtering out the unreliable ones \cite{crowdsourcing-quality,crowdsourcing}. 
	Note that this field is still fairly recent, and the problems mentioned above need to be solved urgently.

	This study intends to develop a novel framework of SLAM with measurement plug-and-play and crowdsourcing capacities.
	We use limited radio features to describe the radio environment and then construct a radio feature map. 
	Our study is inspired by the recently introduced belief propagation (BP) algorithm \cite{BP-SLAM,BP2,slam2}, and 
	we inherit the idea of probabilistic data associations and sequential estimations.
	First, our development is in accordance with the existing BP-based SLAM algorithm \cite{BP2} with the enhancement from allowing different categories of measurements with biases, denoted as measurement plug-and-play mechanism. This mechanism leads to significant improvement in real-world applications.
	Second, a probabilistic crowdsourcing mechanism is proposed to enhance the SLAM performance, thereby enabling multi-agent cooperation.
	Our main contributions are presented in detail as follows:
	\begin{itemize}
		
		\item Various measurements can be obtained through different sensors and RF signals.
		We realize a measurement plug-and-play mechanism by extending the classic BP-based SLAM method.
		We divide the measurements into three categories according to their unknown biases: \textit{agent dependent}, \textit{agent-anchor dependent}, and \textit{agent-feature dependent}. 
		Then, we explain the mechanism in detail by taking TOA, AOA, and RSS measurements as examples, for which the measurements have uncertain values, such as clock bias, orientation bias, and unknown RSS model parameters.	
		We show that these uncertain values can be estimated automatically by the proposed plug-and-play SLAM mechanism.

		\item Numerous agents must cooperate in the construction, exploitation, and refinement of the radio map.	
		We develop a probabilistic crowdsourcing SLAM mechanism to achieve such cooperation and design a corresponding physical layer frame structure. 
		We provide clear definitions of radio feature maps in the local agent and in the cloud.
		We propose an effective solution to the ``double count" and ``data reliability" problems by embedding the probabilistic data association algorithm to the crowdsourcing mechanism, calculating the weighted existence probability of features, and pruning.	
		We reveal that the proposed crowdsourcing mechanism can further improve the performance of the proposed plug-and-play SLAM.

		\item Compared with existing SLAM methods,
		the proposed plug-and-play SLAM with crowdsourcing can work successfully in more realistic and challenging scenarios without prior information about the floor plan, anchors, or agents.
		Moreover, the proposed technique can work under different measurement conditions according to devices' capabilities, thereby making the method suitable for heterogeneous IoT devices.
		The information exchange is facilitated by abstracting the radio environment into the radio feature map, thus reducing data exchange.
		The privacy of each agent can also be protected by only sharing radio feature maps.
		
	\end{itemize}

	The rest of this paper is organized as follows: in Section \uppercase\expandafter{\romannumeral2}, we introduce the system model. The SLAM with measurement plug-and-play and crowdsourcing mechanisms is proposed in Section \uppercase\expandafter{\romannumeral3}. Our simulation results are presented in Section \uppercase\expandafter{\romannumeral4}, and we conclude the paper in Section \uppercase\expandafter{\romannumeral5}.

	\section{System Model}\label{s2}

\begin{figure}
		\centering
		\includegraphics[scale=1.1]{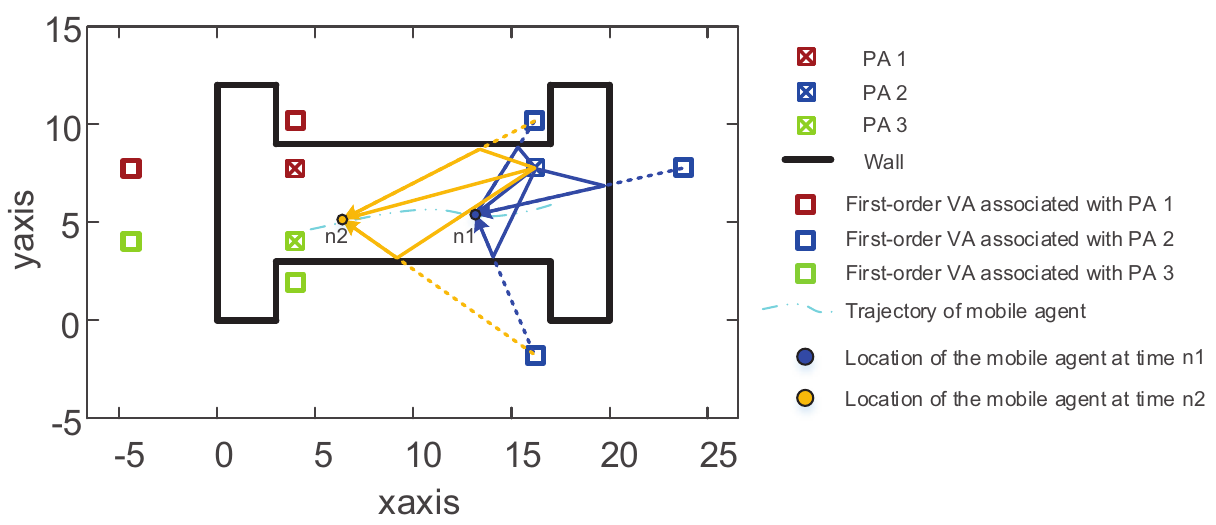}
		\caption{Floor plan with $3$ PAs (denoted by crossed boxes). Each PA has the associated first-order virtual anchors (VAs) (represented by squares), and the  trajectory of a mobile agent is depicted by a blue double-dotted line.} \label{fig:floorplan}
	\end{figure}

	We consider an indoor scenario with $M$ static physical anchors (PAs) and a mobile agent (multi-agent scenarios are considered in Section \ref{cs}).
	An antenna array with $N_R$ elements is equipped at the mobile agent.
	At time slot $n$, the received signal of the mobile agent from the $m$-th PA is given by 
	\begin{equation}\label{receiver}
	\mathbf{y}_{n}^{(m)}\!(t)= 
	\sum_{l=1}^{L_{n}^{(m)}}\alpha^{(m)}_{n,l}\mathbf{a}_{\text{R}}(\theta_{n,l}^{(m)})s^{(m)}(t-\tau_{n,l}^{(m)})+\mathbf{n}^{(m)}(t),
	\end{equation}
	where ${L}_{n}^{(m)}$ is the total number of line-of-sight (LOS) and first-order specular non-line-of-sight (NLOS) propagation paths;
	$\alpha_{n,l}^{(m)}$ is the complex path gain; 
	$\mathbf{a}_{\text{R}}(\cdot) \in \mathbb{C}^{N_R \times 1}$ is the steering vector;
	$\theta_{n,l}^{(m)}$ is the AOA;  
	$\tau_{n,l}^{(m)}$ is the TOA;
	$s^{(m)}(t)$ is the pilot signal, and we assume that the pilot sent by different PAs are orthogonal to each other;
	and $\mathbf{n}^{(m)}(t)\in \mathbb{C}^{N_R \times 1}$ includes the diffuse multipath interfere and the additive white Gaussian noise.
	As shown in Fig. \ref{fig:floorplan}, the mobile agent receives one LOS path and three first-order specular NLOS paths from the PA $2$ at the time slot $n_1$.
	For brevity, we only take PA $2$ as an example for illustration in Fig. \ref{fig:floorplan}.
	
	The geometrical relationship between the mobile agent and PAs can be established by 
	the specular MPCs.
	The state of the mobile agent at time $n$ is denoted as $\mathbf{u}_{n} = [x_{{u}_{n}},y_{{u}_{n}},\dot{x}_{{u}_{n}},\dot{y}_{{u}_{n}}]$, where $[x_{{u}_{n}},y_{{u}_{n}}]$ indicates the location and $[\dot{x}_{{u}_{n}},\dot{y}_{{u}_{n}}]$ is the velocity.
	As the GPS is blocked indoors, the trajectory of the agent is unknown.
	Let $\mathbf{p}^{(m)}_{n,1} = [x_{{p}^{(m)}_{n,1}},y_{{p}^{(m)}_{n,1}}]$
	represent the location of the $m$-th PA at time $n$, for $m=1,\ldots,M$.
	We denote VAs as mirror images of PAs at the reflecting surfaces.
	Therefore, one first-order specular NLOS path corresponds to one VA.
	The location of the $l$-th VA of the $m$-th PA at time $n$ is represented as $\mathbf{p}^{(m)}_{n,l} = [x_{{p}^{(m)}_{n,l}},y_{{p}^{(m)}_{n,l}}]$, where $l=2,\ldots,L_{n}^{(m)}$.
	VAs have two characteristics, that is, a (i) static location and (ii) changing presence.
	The first characteristic is due to the assumption that PAs and scatterers are static.
	The second characteristic arises from the fact that the agent encounters different scatterers while moving.
	As illustrated in Fig. \ref{fig:floorplan}, where we take PA $2$ and the associated VAs as an example, 
	at the time slot $n_1$, the agent observes three VAs associated with PA $2$,
	whereas, at the time slot $n_2$, the agent moves to another location and sees only two VAs associated with PA $2$.
	Although the presence of VAs changed (the number of observed VAs changed from three to two), the location of each VA remained unchanged.
	
	Obtaining the location of the PAs and VAs is useful for communication and localization.
	For example, if a mobile agent knows the location of PAs and VAs, then beam training can be easily accomplished by directing the beams to the PAs and VAs.
	When the location of PAs and VAs is known, multi-point localization can also be realized.
	Therefore, constructing a radio map indicating the location of PAs and VAs in a range of interest (ROI) is important.
	PAs and VAs are regarded as \emph{features} of a radio map. 
	In an unfamiliar environment, the floor plan is typically unknown.
	Therefore, the location of features, that is, $\mathbf{p}^{(m)}_{n,l}$ for $l=1,\ldots,L_{n}^{(m)}$ and $m=1,\ldots,M$, is unknown in advance.
	In addition, given that VAs have a changing presence, $L_{n}^{(m)}$, the total number of features corresponding to the $m$-th PA, is unknown and varies with time.
	We assume that $M$ PAs and $V$ VAs exist in the ROI, and thus there are $2(M+V)$ unknowns, because the location of each feature is described by 2-dimensional (2-D) coordinates.
	Estimating $2(M+V)$ unknowns with the measurements obtained by one agent at one time slot is impossible.
	Nevertheless, as the locations of PAs and VAs are static, we have a chance to obtain their estimates with a large number of measurements accumulated over time.
	
	\begin{table*}
		\centering
		\caption{Notations of important variables.}\label{NOTATIONS}
		\renewcommand{\arraystretch}{1.5}
		\fontsize{7.9}{7.9}\selectfont
		\begin{threeparttable}	
			\begin{tabular}{c l c l}			
				\toprule		
				Notation & Definition & Notation & Definition \\
				\hline
				$\mathbf{p}^{(m)}_{n,l}$ & $l=1$: location of the $m$-th PA at time $n$                  & $\mathbf{z}_{n,l}^{(m)}$ & measurement vector corresponding to the $l$-th VA \\  
				& $l>1$: location of the $l$-th VA of the $m$-th PA at time $n$ &                          & of the $m$-th PA at time $n$                      \\ \hline
				$\mathbf{u}_{n}$         & state of the mobile agent at time $n$                         & $\mathbf{a}_n^{(m)}$     & feature-oriented data association vector          \\ \hline
				$L_{n}^{(m)}$            & number of features corresponding to the $m$-th PA             & $\mathbf{b}_n^{(m)}$     & measurement-oriented data association vector      \\ \hline
				$\mathbf{s}^{(m)}_{n,l}$ & unknown measurement biases associated              & ${\mathbf{v}}^{(m)}_{n,l}$ & state of the $l$-th feature of the $m$-th PA at time $n$                                   \\ 
				& with the $l$-th feature of the $m$-th PA at time $n$            &                            & ${\mathbf{v}}^{(m)}_{n,l} = [\mathbf{p}^{(m)}_{n,l},\mathbf{s}^{(m)}_{n,l},r^{(m)}_{n,l}]$ \\ \hline
				$r^{(m)}_{n,l}$          & binary variable indicates the existence of feature & $\mathbf{c}_{n}$           & number-of-measurements vector at time $n$                                                 \\ \hline
				$\alpha_n$          & unknown orientation bias of the agent              & $\omega_n^{(m)}$  & unknown clock bias between agent and the $m$-th PA  \\ \hline
				$\beta_{n,l}^{(m)}$ & path loss exponent of the $l$-th VA of the $m$-th PA            & $\xi_{n,l}^{(m)}$ & reference RSS of the $l$-th VA of the $m$-th PA \\ \hline
				$\mathcal{K}_{n-1}^{(m)}$     & set of legacy features from time slot $n-1$ & $\mathcal{M}_{n}^{(m)}$ & set of obtained MPC measurement indexes of the $m$-th PA                                                            \\ \hline
				$\mathcal{D}_{n}^{(m)}$ & set of legacy feature indexes which generate measurement     & $\bar{\mathcal{D}}_{n}^{(m)}$ & $\bar{\mathcal{D}}_{n}^{(m)}= \mathcal{K}_{n-1}^{(m)} \backslash {\mathcal{D}}_{n}^{(m)}$                     \\ \hline
				$\mathcal{N}_{n}^{(m)}$ & set of measurement indexes originate from new features           & $\bar{\mathcal{N}}_{n}^{(m)}$ & $\bar{\mathcal{N}}_{n}^{(m)}= \mathcal{M}_{n}^{(m)} \backslash {\mathcal{N}}_{n}^{(m)}$                       \\ \hline
				$\mathcal{F}_{n}^{(m)}$ & set of measurement indexes of false alarms                       & $\mathcal{L}_k$               & LRF-Map of the $k$-th agent, $\mathcal{L}_k = \Big\{ \mathcal{L}_k^{(1)}, \ldots, \mathcal{L}_k^{(M)} \Big\}$ \\ \hline
				$\mathcal{O}$           & ORF-Map, $\mathcal{O} = \Big\{ \mathcal{O}^{(1)}, \ldots, \mathcal{O}^{(M)} \Big\}$ & $\varphi_{k,n'}$              & weight coefficient of the $k$-th agent at time $n'$                                         \\ \hline
				$\tilde{\ast}$ & legacy $\ast$, $\ast = \{   \mathbf{p}^{(m)}_{n,l}, \mathbf{s}^{(m)}_{n,l}, r^{(m)}_{n,l}, {\mathbf{v}}^{(m)}_{n,l}, \beta_{n,l}^{(m)}, \xi_{n,l}^{(m)}\}$  & $ \breve{\star}$ & new $\star$, $\star = \{   \mathbf{p}^{(m)}_{n,l}, \mathbf{s}^{(m)}_{n,l}, r^{(m)}_{n,l}, {\mathbf{v}}^{(m)}_{n,l}, \beta_{n,l}^{(m)}, \xi_{n,l}^{(m)}\}$  \\				
				\bottomrule			
			\end{tabular}\vspace{-0.4cm}
		\end{threeparttable}
	\end{table*}
	
	Let $\mathcal{M}_{n}^{(m)}$ represent a set of obtained MPC measurement indexes of the $m$-th PA at time $n$.
	The MPC measurements, including the extracted RSS $\lvert\hat{\alpha}^{(m)}_{n,l}\lvert$, TOA $\hat{\tau}_{n,l}^{(m)}$, and AOA $\hat{\theta}_{n,l}^{(m)}$, where $l = 1,\ldots,|\mathcal{M}_{n}^{(m)}|$, can be effectively obtained from the received signal $\mathbf{y}_{n}^{(m)}(t)$ by advanced channel parameters extraction algorithms \cite{hy,nomp,yj1,yj2}.
	Note that $|\cdot|$ denotes the amplitude of a complex value or the number of elements in the set.
	Moreover, let $\mathbf{z}_{n,l}^{(m)}$ denote a vector of measurements of the $l$-th MPC corresponding to the $m$-th PA at time $n$, e.g., $\mathbf{z}_{n,l}^{(m)}=[\hat{\alpha}^{(m)}_{n,l},\hat{\tau}_{n,l}^{(m)}, \hat{\theta}_{n,l}^{(m)}]$.
	We can then define the stacked measurement vectors
	$\mathbf{z}_{n}^{(m)}=[\mathbf{z}_{n,1}^{(m)},\cdots,\mathbf{z}^{(m)}_{n,|\mathcal{M}_{n}^{(m)}|}]$ and
	$\mathbf{z}_{n}=[\mathbf{z}_{n}^{(1)},\ldots,\mathbf{z}_{n}^{(M)}]$.
	Through the accumulation of $N$ time slots, the agent can obtain a sequence of measurements $\mathbf{z}_{1:N}=[\mathbf{z}_{1},\ldots,\mathbf{z}_{N}]$.

	\section{SLAM with Measurement Plug-and-Play and Crowdsourcing Mechanisms}
	We consider a challenging scenario in which an agent enters an unfamiliar indoor environment (e.g., a large shopping mall or underground garage), where the conventional GPS is blocked.
	Therefore, $\mathbf{u}_{n}$ and $\mathbf{p}^{(m)}_{n,l}$ are unknown for $l=1,\ldots,L_{n}^{(m)}$, $m=1,\ldots,M$, and $n=1,\ldots,N$. 
	Only the location of the entrance may be known and provides an approximate starting point for the agent.
	The aim of this study is to estimate the agent's state $\mathbf{u}_{n}$ and the features' locations  $\mathbf{p}^{(m)}_{n,l}$ for $l=1,\ldots,L_{n}^{(m)}$, $m=1,\ldots,M$, and $n=1,\ldots,N$ with the obtained sequence of measurements $\mathbf{z}_{1:N}=[\mathbf{z}_{1},\ldots,\mathbf{z}_{N}]$.
	We establish plug-and-play and crowdsourcing mechanisms for  SLAM to realize such a goal.
	The important variables are summarized in Table \ref{NOTATIONS}.

	\subsection{Theoretical Foundation for BP-based SLAM}\label{TF}
	Given the data association uncertainty, 
	a measurement can originate from a \emph{legacy feature}, \emph{new feature}, or did not originate from any feature (i.e., a \emph{false alarm}).
	A legacy feature means the feature already exists at time $n\!-\!1$.
	Let $\tilde{\mathbf{p}}^{(m)}_{n,l}$ denote the location of a legacy feature, and $\mathcal{D}_{n}^{(m)}$ denote the set of legacy feature indexes which generate measurement at time $n$.
	By contrast, a new feature means the feature does not exist at time $n\!-\!1$.
	Let $\breve{\mathbf{p}}^{(m)}_{n,l}$ denote the location of a new feature, and $\mathcal{N}_{n}^{(m)}$ denote the set of measurement indexes originate from new features.
	Finally, let $\mathcal{F}_{n}^{(m)}$ denote the set of measurement indexes of false alarms. Therefore, we classify the measurement indexes in $\mathcal{M}_{n}^{(m)}$ into three subsets according to their origins and obtain
	$
	|\mathcal{M}_{n}^{(m)}| = |\mathcal{D}_{n}^{(m)}|+|\mathcal{N}_{n}^{(m)}|+|\mathcal{F}_{n}^{(m)}|
	$.

	To associate measurements with features, we define the following two data association vectors according to \cite{BP1,BP2}.
	First, let $\mathcal{K}_{n-1}^{(m)}$ represent a set of legacy features indexes from time slot $n-1$.
	The $|\mathcal{K}_{n-1}^{(m)}|$-dimensional feature-oriented vector is  $\mathbf{a}_n^{(m)}=\big[{a}_{n,1}^{(m)},\ldots,{a}_{n,|\mathcal{K}_{n-1}^{(m)}|}^{(m)}\big]$, the element of which are given by 
	\begin{equation}\label{DA1}
	{a}_{n,i}^{(m)}\!=\!\left\{
	\begin{array}{ll}
	\!\!\!j \in \mathcal{M}_{n}^{(m)}, &\!\!\text{legacy feature $i$ of PA $m$ generates measurement $j$ at time $n$,} \\
	\!\!\!0, &\!\!\text{legacy feature $i$ of PA $m$ does not generate any measurement at time $n$},
	\end{array} \right.
	\end{equation}where $i = 1,\ldots, |\mathcal{K}_{n-1}^{(m)}|$.
	We also define the stacked vector $\mathbf{a}_n=[\mathbf{a}_{n}^{(1)},\ldots,\mathbf{a}_{n}^{(M)}]$.
	Second, the $|\mathcal{M}_{n}^{(m)}|$-dimensional measurement-oriented vector is $\mathbf{b}_n^{(m)}=\big[{b}_{n,1}^{(m)},\ldots,{b}_{n,|\mathcal{M}_{n}^{(m)}|}^{(m)}\big]$, where we obtain 
	\begin{equation}\label{DA2}
	{b}_{n,j}^{(m)}\!=\!\left\{
	\begin{array}{ll}
	\!\!\! i \in \mathcal{K}_{n-1}^{(m)}, & \!\!\text{measurement $j$ is generated by legacy feature $i$ of PA $m$ at time $n$,} \\ 
	\!\!\! 0, &\!\!\text{measurement $j$ is not generated by any legacy feature at time $n$,} 
	\end{array} \right.
	\end{equation}
	for $j = 1,\ldots, |\mathcal{M}_{n}^{(m)}|$.
	We also define the stacked vector $\mathbf{b}_n=[\mathbf{b}_{n}^{(1)},\ldots,\mathbf{b}_{n}^{(M)}]$.
	The two vectors $\mathbf{a}_n$ and $\mathbf{b}_n$, which are equivalent, as one can be determined from the other,
	can ensure the scalability properties of the BP algorithm. 
	A constraint exists such that each measurement originates from a maximum of one feature or one false alarm, and one feature can generate at most one measurement each time.
	The exclusion-enforcing
	function used to ensure the constraint is defined as
	\begin{equation}\label{indicator}
	\Psi(\mathbf{a}_{n}^{(m)},\mathbf{b}_{n}^{(m)})= \prod\limits_{i=1}^{|\mathcal{M}_{n}^{(m)}|}\prod\limits_{j=1}^{|\mathcal{K}_{n-1}^{(m)}|}\Psi({a}_{n,i}^{(m)},{b}_{n,j}^{(m)}),
	\end{equation}
	where 
	\begin{equation}
	\Psi({a}_{n,i}^{(m)},{b}_{n,j}^{(m)}) = \left\{
	\begin{array}{ll}
	0,  &  {a}_{n,i}^{(m)}=j, {b}_{n,j}^{(m)} \neq i\  \text{or} \  {b}_{n,j}^{(m)}=i, {a}_{n,i}^{(m)} \neq j,\\
	1, &  \text{otherwise}.
	\end{array} \right.
	\end{equation}
	
	We define the likelihood function, that is, the distribution of measurements conditional on the unknown state of agent and features, and two data association vectors as 
	\begin{equation}\label{mmp}
	f(\mathbf{z}^{(m)}_{n}|\mathbf{u}_{n},{\mathbf{v}}_{n}^{(m)},\mathbf{a}_{n}^{(m)},\mathbf{b}_{n}^{(m)}) 
	=\!\!\!\! \prod\limits_{i\in\mathcal{D}_{n}^{(m)}}\!\!  \!f(\mathbf{z}^{(m)}_{n,{a}_{n,i}^{(m)}}|\mathbf{u}_{n},\tilde{{\mathbf{v}}}_{n}^{(m)})
	 \!\!\prod\limits_{ j\in\mathcal{N}_{n}^{(m)}}\!\!\!f(\mathbf{z}^{(m)}_{n,j}|\mathbf{u}_{n},\breve{{\mathbf{v}}}_{n}^{(m)}) 
\!\! \prod\limits_{k\in\mathcal{F}_{n}^{(m)}}\!\!\!f_{\rm false}(\mathbf{z}^{(m)}_{n,k}),
	\end{equation}
	where $\mathcal{D}_{n}^{(m)} \triangleq\left\{i \in\{1, \ldots, |\mathcal{K}_{n-1}^{(m)}|\}: a_{n, i}^{(m)} \neq 0\right\}$;
	$\tilde{{\mathbf{v}}}_{n}^{(m)}$ and $\breve{{\mathbf{v}}}_{n}^{(m)}$ denote the state of legacy and new features, respectively; and 
	${\mathbf{v}}_{n}^{(m)}=[\tilde{{\mathbf{v}}}_{n}^{(m)},\breve{{\mathbf{v}}}_{n}^{(m)}]$.
	The elements in ${\mathbf{v}}_{n}^{(m)}$ are defined as ${\mathbf{v}}^{(m)}_{n,l} = [\mathbf{p}^{(m)}_{n,l},\mathbf{s}^{(m)}_{n,l},r^{(m)}_{n,l}]$ for $l=1,\ldots, |\mathcal{D}_{n-1}^{(m)}|+|\mathcal{N}_{n}^{(m)}|$,
	where $\mathbf{s}^{(m)}_{n,l}$ represents unknown measurement biases associated with the $(m,l)$-th feature at time $n$, and a binary variable
	$r^{(m)}_{n,l}\in\{0,1\}$ indicates the existence of the $(m,l)$-th feature at time $n$, that is, the feature exists at time $n$ if and only if $r^{(m)}_{n,l}=1$. Similarly, the elements in $\tilde{{\mathbf{v}}}_{n}^{(m)}$ and $\breve{{\mathbf{v}}}_{n}^{(m)}$ are denoted as $\tilde{\mathbf{v}}^{(m)}_{n,l} = [\tilde{\mathbf{p}}^{(m)}_{n,l},\tilde{\mathbf{s}}^{(m)}_{n,l},\tilde{r}^{(m)}_{n,l}]$ for $l=1,\ldots, |\mathcal{D}_{n-1}^{(m)}|$ and $\breve{\mathbf{v}}^{(m)}_{n,l} = [\breve{\mathbf{p}}^{(m)}_{n,l},\breve{\mathbf{s}}^{(m)}_{n,l},\breve{r}^{(m)}_{n,l}]$ for $l=1,\ldots,|\mathcal{N}_{n}^{(m)}|$, respectively.
	The number of false alarms and newly detected features follows a Poisson distribution with a mean of $\mu_{\rm false}^{(m)}$ and $\mu_{\rm new}^{(m)}$, respectively.
	The distribution of each false alarm measurement is described by the probability density function (pdf) $f_{\rm false}(\cdot)$.

\begin{figure}[t]
		\centering
		\includegraphics[scale=0.6]{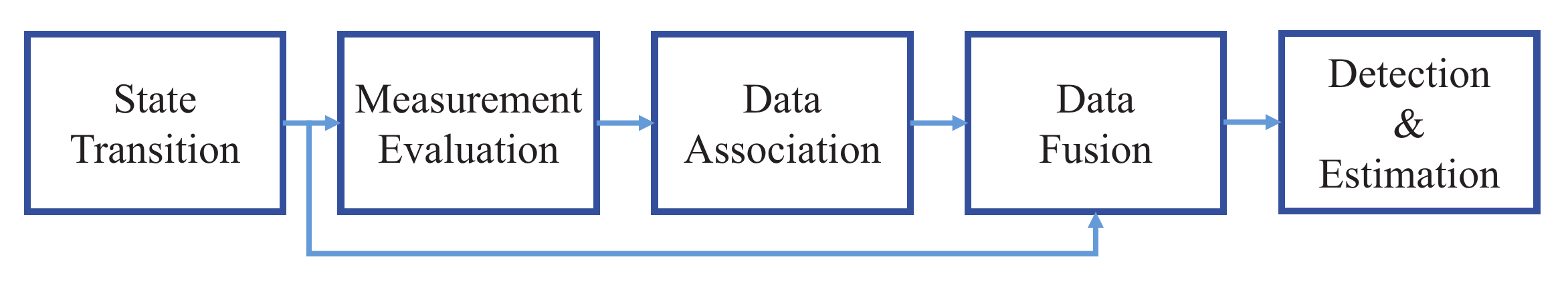}
		\caption{Framework of a classic BP-based SLAM algorithm.} \label{fig:as}	
	\end{figure}

	The joint posterior probability distribution of the state of the agent and features and the data association vectors conditioned on measurements for all time slots up to $N$ is defined as
	\begin{equation}\label{jointP}
	f(\mathbf{u}_{1:N},{\mathbf{v}}_{1:N},\mathbf{a}_{1:N},\mathbf{b}_{1:N}|\mathbf{z}_{1:N})
	= \prod\limits_{n=1}^{N} \prod\limits_{m=1}^{M} f(\mathbf{u}_{n},{\mathbf{v}}_{n}^{(m)},\mathbf{a}_{n}^{(m)},\mathbf{b}_{n}^{(m)}|\mathbf{z}_{n}^{(m)}),
    \end{equation}
	which 
	can then be computed according to Bayes' theorem as
	\begin{multline}\label{jointP1}
 f(\mathbf{u}_{1:N},{\mathbf{v}}_{1:N},\mathbf{a}_{1:N},\mathbf{b}_{1:N}|\mathbf{z}_{1:N})
	\!\propto \! \prod\limits_{n=1}^{N}\!\prod\limits_{m=1}^{M}\!\underbrace {f(\mathbf{u}_{n},\tilde{{\mathbf{v}}}_{n}^{(m)}|\mathbf{u}_{n-1},{\mathbf{v}}_{n-1}^{(m)})}_{(a)}
	\underbrace {f(\mathbf{z}_{n}^{(m)}|\mathbf{u}_{n},{\mathbf{v}}_{n}^{(m)},\mathbf{a}_{n}^{(m)},\mathbf{b}_{n}^{(m)})}_{(b)}
	\\
	 \times\underbrace {f(\mathbf{a}_{n}^{(m)},\mathbf{b}_{n}^{(m)},{c}^{(m)}_{n},\breve{{\mathbf{v}}}_{n}^{(m)}|\tilde{{\mathbf{v}}}_{n}^{(m)},\mathbf{u}_{n})}_{(c)},
	\end{multline}
	where ${c}_{n}^{(m)}=|\mathcal{M}_{n}^{(m)}|$ and $\mathbf{c}_n=[{c}_{n}^{(1)},\ldots,{c}_{n}^{(M)}]$ is the number-of-measurements vector at time $n$, and we have $\mathbf{a}_{n}$ implies $\mathbf{b}_{n}$ and $\mathbf{z}_n$ implies $\mathbf{c}_{n}$.
	The framework of a classic BP-based SLAM algorithm is derived from \eqref{jointP1}. 
	Note that (a), (b), and (c) of \eqref{jointP1} correspond to the state transition, measurement evaluation, and data association phases, respectively. The entire process of \eqref{jointP1} realizes the data fusion phase (Fig. \ref{fig:as}).

	Given the posterior probability, a minimum mean square error (MMSE) estimator for the agent's state $\mathbf{u}_{N}$ at time slot $N$ is given as  
	\begin{equation}\label{mmse1}
	\hat{\mathbf{u}}_{N} = \int \mathbf{u}_{N} f(\mathbf{u}_{N}|\mathbf{z}_{1:N})  \text{d} \mathbf{u}_{N}, 
	\end{equation}
	where $f(\mathbf{u}_{N}|\mathbf{z}_{1:N})= \int_{\mathbf{w}}  f(\!\mathbf{u}_{1:N},\!{\mathbf{v}}_{1:N},\!\mathbf{a}_{1:N},\!\mathbf{b}_{1:N}\!|\mathbf{z}_{1:N}\!) \text{d}\mathbf{w} $  is a marginal posterior distribution in \eqref{jointP1}, and $\mathbf{w} = [\mathbf{u}_{1:N\!-\!1},\!{\mathbf{v}}_{1:N},\!\mathbf{a}_{1:N},\!\mathbf{b}_{1:N}]$.
	The posterior existence probability $p({r}_{N,l}^{(m)}=1|\mathbf{z}_{1:N})$ is given as
	\begin{equation}\label{marginal}
	p({r}_{N,l}^{(m)}\!=\!1|\mathbf{z}_{1:N}) \!=\!\!\! \iint \!\!\!f(\mathbf{{p}}_{N,l}^{(m)},\!\mathbf{{s}}_{N,l}^{(m)},\!{r}_{N,l}^{(m)}\!=\!1|\mathbf{z}_{1:N}) \text{d} \mathbf{p}_{N,l}^{(m)} \text{d} \mathbf{s}_{N,l}^{(m)},
	\end{equation} 
	where $f(\mathbf{{p}}_{N,l}^{(m)},\!\mathbf{{s}}_{N,l}^{(m)},\!{r}_{N,l}^{(m)}\!=\!1|\mathbf{z}_{1:N})$ is a marginal posterior distribution in \eqref{jointP1}.
	According to Bayes' theorem, we obtain 
	\begin{equation}\label{marginal2}
	f(\mathbf{p}_{N,l}^{(m)}|{r}_{N,l}^{(m)}\!=\!1,\mathbf{z}_{1:N})  \!=\! \frac{ \int\!\! f(\mathbf{p}_{N,l}^{(m)}\!,\mathbf{{s}}_{N,l}^{(m)}\!,{r}_{N,l}^{(m)}\!=\!1|\mathbf{z}_{1:N})\text{d} \mathbf{s}_{N,l}^{(m)}}{p({r}_{N,l}^{(m)}=1|\mathbf{z}_{1:N})}. 
	\end{equation}
	The MMSE estimator for the feature location $\mathbf{p}_{N,l}^{(m)}$ can be obtained similarly as 
	\begin{equation}\label{mmse6}
	\hat{\mathbf{p}}_{N,l}^{(m)} \! =\!\! \int \!\! \mathbf{p}_{N,l}^{(m)} f(\mathbf{p}_{N,l}^{(m)}|{r}_{N,l}^{(m)}=1,\mathbf{z}_{1:N})  \text{d} \mathbf{p}_{N,l}^{(m)}. 
	\end{equation}
	Direct marginalization from \eqref{jointP1} is infeasible.
	Hence,
	the BP algorithm can be applied to approximate marginal posterior pdfs \cite{sumpro}.
	The detection phase in Fig. \ref{fig:as} is according to \eqref{marginal}, and the estimation phase is according to \eqref{mmse1} and \eqref{mmse6}.

	\subsection{Measurement Plug-and-Play Mechanism}\label{mpp}
	In this subsection, we establish a flexible fusion mechanism in which measurements can be plugged in and played.
	First, we classify the measurements into three categories according to the unknown biases as
	\textit{agent dependent}, \textit{agent-anchor dependent}, and \textit{agent-feature dependent}. 
	The agent-dependent measurements represent the unknown bias of measurements dependent only on the agent device. Measurements from the built-in sensors of the agent device, such as a gyro or accelerator, are agent-dependent measurements. 
	In this study, we take measurements extracted from RF signals as examples. 
	The AOA measurements at the agent side are agent-dependent measurements, and the relationship between the AOA measurements and agent and feature locations is given as  
	\begin{equation}\label{AOA}
	\text{AOA}(\mathbf{u}_{n},\mathbf{p}^{(m)}_{n,l})\!=\!\arctan\!\left(\!\dfrac{y_{p_{n,l}^{(m)}}\!-\!y_{u_n}}{x_{p_{n,l}^{(m)}}\!-\!x_{u_n}}\!\right)\!+\!\alpha_n,
	\end{equation}
	where $\alpha_n$ is the unknown orientation bias of the agent. 
	The agent-anchor dependent measurements stand for the unknown bias of the measurements and are dependent on different agent-anchor pairs (the anchor here means a PA). For instance, 
	the relationship between TOA measurements and agent and feature locations is given as 
	\begin{equation}\label{TOA}
	\text{TOA}(\mathbf{u}_{n},\mathbf{p}^{(m)}_{n,l})=\dfrac{\big\lVert \mathbf{u}_{n}[1:2]-\mathbf{p}^{(m)}_{n,l} \big\lVert-\omega_n^{(m)}}{c},
	\end{equation}
	where $\|\cdot\|$ signifies the L$_2$-norm, $c$ is the speed of light, and $\omega_n^{(m)}$ is the unknown clock bias that depends on different agent-anchor pairs.
	The agent-feature dependent measurements mean that the unknown bias of the measurements is dependent on different agent-feature pairs, such as RSS measurements. 
	In practice, each anchor may use different transmission powers, and the surrounding environment of each feature may differ. Therefore, the RSS model parameters of each agent-feature pair vary and are unknown. 
	To relate the determined RSS values to the agent and feature locations, the path loss model \cite{RSS} is given by
	\begin{equation}\label{RSS}
	\text{RSS}(\mathbf{u}_{n},\mathbf{p}^{(m)}_{n,l})=-10\beta_{n,l}^{(m)}\log_{10}\big\lVert \mathbf{u}_{n}[1:2]-\mathbf{p}^{(m)}_{n,l} \big\lVert+\xi_{n,l}^{(m)},
	\end{equation}
	where $\beta_{n,l}^{(m)}$ is the path loss exponent (PLE), which varies depending on the environment, and $\xi_{n,l}^{(m)}$ is a fixed constant that accounts for the transmit power (called reference RSS, RefRSS).
	The unknown parameters $\beta_{n,l}^{(m)}$ and $\xi_{n,l}^{(m)}$ correspond to different agent-feature pairs.
	Despite taking AOA, TOA, and RSS as examples, they merely represent three different categories of measurements, and measurements are not limited to the RF signals.
	
	\begin{remark}	
		Although a subscript $n$ exists in $\alpha_n$,  $\omega_n^{(m)}$, $\beta_{n,l}^{(m)}$, and $\xi_{n,l}^{(m)}$, these parameters are time-invariant, similar to the location of features $\mathbf{p}^{(m)}_{n,l}$.
		The orientation of the mobile agent changes, but the relative orientation is recorded by the built-in gyro. For simplicity, we regard the relative orientation as zero and denote $\alpha_n$ in \eqref{AOA} as the orientation bias.	
	\end{remark}
	
	Different categories of measurement are integrated into the SLAM system in distinct ways.
	The statistical formulas presented in Section \ref{TF} undergo the following changes.
	According to the previously defined vector ${\mathbf{v}}^{(m)}_{n,l} = [\mathbf{p}^{(m)}_{n,l},\mathbf{s}^{(m)}_{n,l},r^{(m)}_{n,l}]$, where $\mathbf{s}^{(m)}_{n,l}$ is a vector of unknown biases.
	Taking AOA, TOA, and RSS measurements as examples,
	we obtain $\mathbf{s}^{(m)}_{n,l}=[\alpha_n, {\omega}_n^{(m)}, \xi_{n,l}^{(m)}, \beta_{n,l}^{(m)}]$.
	As $\alpha_n$ and ${\omega}_n^{(m)}$ are irrelevant to the state of VAs, we remove these two elements from $\mathbf{s}^{(m)}_{n,l}$.
	Therefore, we obtain $\mathbf{s}^{(m)}_{n,l} = [\xi_{n,l}^{(m)},\beta_{n,l}^{(m)}]$ and the vector 
	${\mathbf{v}}^{(m)}_{n,l} = [\mathbf{{p}}^{(m)}_{n,l},\xi_{n,l}^{(m)},\beta_{n,l}^{(m)},{r}^{(m)}_{n,l}]$. 
	Let $\tilde{\xi}_{n,l}^{(m)}$ and $\tilde{\beta}_{n,l}^{(m)}$ denote parameters corresponding to legacy features and $\breve{\xi}_{n,l}^{(m)}$ and $\breve{\beta}_{n,l}^{(m)}$ represent parameters corresponding to new features.
	We denote stacked vectors as $\bm{\alpha}_{1:N} \!\!=\!\! [{\alpha}_1,\ldots,{\alpha}_N]$ and
	$\bm{\omega}_{1:N}\!\!=\!\![\bm{\omega}_1,\ldots,\bm{\omega}_N]$, where
	$\bm{\omega}_n = [{\omega}_n^{(1)},\ldots,{\omega}_n^{(M)}]$.
	
	\subsubsection{State Transition}\label{st}
	The agent state $\mathbf{u}_{n}$ and legacy feature state ${\tilde{\mathbf{v}}}^{(m)}_{n,l}$ are assumed to independently evolve according to Markovian state dynamics given by 
	\begin{equation}\label{mark}
	f(\mathbf{u}_{n},\!{\alpha}_n,\!\bm{\omega}_n,\!\tilde{{\mathbf{v}}}_{n}|\mathbf{u}_{n-1},\!{\alpha}_{n-1},\!\bm{\omega}_{n-1},\!{\mathbf{v}}_{n-1})\!\!=\!\!\!f(\mathbf{u}_{n}\!|\mathbf{u}_{n-1})f(\alpha_{n}\!|\alpha_{n-1}) \!\!\prod\limits_{m=1}^{M}\!\!\!f({\omega}_n^{(m)}|{\omega}_{n-1}^{(m)})\!\!\!\!\prod\limits_{l=1}^{|\mathcal{K}_{n-1}^{(m)}|}\!\!\!\!f(\tilde{{\mathbf{v}}}^{(m)}_{n,l}|{\mathbf{v}}^{(m)}_{n-1,l}).
	\end{equation}
	The state transition function of agent $f(\mathbf{u}_{n}|\mathbf{u}_{n-1})$
	is defined by a linear, near constant-velocity motion model \cite{sem}, given as $\mathbf{u}_{n}^{\rm T} = \mathbf{A} \mathbf{u}_{n-1}^{\rm T}+ \mathbf{d}_{n}$, where $(\cdot)^{\text{T}}$ represents the transpose,
	and 
	\begin{equation}       
	\mathbf{A}=\left(                
	\begin{array}{cccc}   
	1 & 0 & \Delta T & 0\\ 
	0 & 1 & 0 & \Delta T\\  
	0 & 0 & 1 & 0\\  
	0 & 0 & 0 & 1\\  
	\end{array}
	\right),
	\end{equation}
	where $\Delta T$ is the sampling period, and
	$\mathbf{d}_{n}$ is the driving process, which follows the independently identically Gaussian distribution across $n$ with zero-mean and covariance matrix $\sigma_d^2 \mathbf{I}$.
	We express
	the state transition function of feature as
	$f(\tilde{{\mathbf{p}}}^{(m)}_{n,l},\tilde{\xi}_{n,l}^{(m)},\tilde{\beta}_{n,l}^{(m)},{\tilde{{r}}}^{(m)}_{n,l}|\mathbf{{p}}^{(m)}_{n-1,l},\xi_{n-1,l}^{(m)},{\beta}_{n-1,l}^{(m)},{\tilde{{r}}}^{(m)}_{n-1,l})$.
	If a feature does not exist at time slot $n-1$, then it cannot exist as a legacy feature at time slot $n$. Therefore,
	for ${\tilde{{r}}}^{(m)}_{n-1,l}=0$, we obtain
	\begin{equation}\label{tt1}
	\hspace{-0.6cm}
	\begin{array}{ll}
	&\!\!\!\!\!f(\tilde{{\mathbf{p}}}^{(m)}_{n,l},\tilde{\xi}_{n,l}^{(m)},\tilde{\beta}_{n,l}^{(m)},{\tilde{{r}}}^{(m)}_{n,l}|\mathbf{{p}}^{(m)}_{n-1,l},\xi_{n-1,l}^{(m)},{\beta}_{n-1,l}^{(m)},0)\!=\!\!\left\{
	\begin{array}{lcc}
	\!\!\!\!f_{\rm D}(\tilde{{\mathbf{p}}}^{(m)}_{n,l},\tilde{\xi}_{n,l}^{(m)},\tilde{\beta}_{n,l}^{(m)}), & & {\tilde{{r}}^{(m)}_{n,l}=0},\\
	\!\!\!\!0, & & {\tilde{{r}}^{(m)}_{n,l}=1},
	\end{array} \right.
	\end{array} 
	\end{equation}
	where $f_{\rm D}(\cdot)$ is an arbitrary ``dummy pdf", which is explained in detail in \cite{BP1} to ensure that $f(\tilde{{\mathbf{v}}}^{(m)}_{n,l}|{\mathbf{v}}^{(m)}_{n-1,l})$ integrates to one.
	If a feature exists at time slot $n-1$, then the probability that it still exists at time slot $n$ is determined by  the survival probability and the state of the feature is distributed according to the state transition pdf.
	Therefore, for ${\tilde{{r}}}^{(m)}_{n-1,l}=1$, we obtain
	\begin{equation}\label{tt2}
	\begin{array}{ll}
	&\!\!\!\!\!f(\tilde{{\mathbf{p}}}^{(m)}_{n,l},\tilde{\xi}_{n,l}^{(m)},\tilde{\beta}_{n,l}^{(m)},{\tilde{{r}}}^{(m)}_{n,l}|\mathbf{{p}}^{(m)}_{n-1,l},\xi_{n-1,l}^{(m)},{\beta}_{n-1,l}^{(m)},1)
	\!=\!\!\left\{
	\begin{array}{lc}
	\!\!\!\!\left(\!1\!-\!P_{\rm s}(\mathbf{{p}}^{(m)}_{n-1,l})\!\right)\!\!f_{\rm D}(\tilde{{\mathbf{p}}}^{(m)}_{n,l},\tilde{\xi}_{n,l}^{(m)},\tilde{\beta}_{n,l}^{(m)}),  & \!\!\!\! {\tilde{{r}}^{(m)}_{n,l}=0},\\
	\!\!\!\!{P_{\rm s}(\mathbf{{p}}^{(m)}_{n-1,l})} f(\tilde{{\mathbf{p}}}^{(m)}_{n,l}|\mathbf{{p}}^{(m)}_{n-1,l})  & \!\!\!\! {\tilde{{r}}^{(m)}_{n,l}=1},\\
	\!\!\!\!\times f({\tilde{\xi}}^{(m)}_{n,l}|{{\xi}}^{(m)}_{n-1,l})f({\tilde{\beta}}^{(m)}_{n,l}|{{\beta}}^{(m)}_{n-1,l}), &
	\end{array} \right.
	\end{array} 
	\end{equation}
	where ${P_{\rm s}(\cdot)}\in (0,1]$ represents the survival probability of a feature.
	
	\begin{remark}
		State transition functions \eqref{tt1} and \eqref{tt2} imply that only the existence probability of features should be updated.	
		As
		${\alpha}_n$, ${\omega}_n^{(m)}$, $\beta_{n,l}^{(m)}$, $\xi_{n,l}^{(m)}$, and ${{\mathbf{p}}}^{(m)}_{n,l}$ are time-invariant, the state transition pdfs are given by
		$f(\ast|\star)=\delta(\ast-\star)$, where $\delta(\cdot)$ is a Dirac delta function; $\ast=\{{\alpha}_n$, ${\omega}_n^{(m)}$, $\tilde{\beta}_{n,l}^{(m)}$, $\tilde{\xi}_{n,l}^{(m)}$, ${\tilde{\mathbf{p}}}^{(m)}_{n,l}\}$; and $\star=\{{\alpha}_{n-1}$, ${\omega}_{n-1}^{(m)}$, $\beta_{n-1,l}^{(m)}$, $\xi_{n-1,l}^{(m)}$, ${{\mathbf{p}}}^{(m)}_{n-1,l}\}$.
		For initialization, we obtain $f(\ast_{1}|\ast_{0})=f(\ast_{1})$, where $\ast=\{\mathbf{u},{\alpha}$, ${\omega}^{(m)}\}$. 
		As ${\mathbf{v}}^{(m)}_{0,l}$ is empty, $\tilde{{\mathbf{v}}}^{(m)}_{1,l}$ is empty as well,
		and we obtain 
		$f(\tilde{{\mathbf{v}}}^{(m)}_{1,l}|{\mathbf{v}}^{(m)}_{0,l})=1$.
	\end{remark}
	
	\subsubsection{Measurement Evaluation}
	The measurements are divided into three subsets according to their origins, as described in \eqref{mmp}.
	The likelihood function is updated to  
	\begin{multline}\label{mp1}
	\!\!\!\!\!\!f(\mathbf{z}^{(m)}_{n}|\mathbf{u}_{n},{\alpha}_{n},\bm{\omega}_{n}^{(m)},{\mathbf{v}}_{n}^{(m)},\mathbf{a}_{n}^{(m)}, {c}_{n}^{(m)})
	\!=\!\!\!\! \prod\limits_{i=1}^{|\mathcal{M}_{n}^{(m)}|} f_{\rm false}(\mathbf{z}^{(m)}_{n,i})  \prod\limits_{j\in\mathcal{D}_{n}^{(m)}} \!\! \dfrac{f(\mathbf{z}^{(m)}_{n,{a}_{n,j}^{(m)}}|\mathbf{u}_{n},{\alpha}_{n},\bm{\omega}_{n}^{(m)},\tilde{{\mathbf{v}}}_{n}^{(m)})}{f_{\rm false}(\mathbf{z}^{(m)}_{n,{a}_{n,j}^{(m)}})} \\
	\times\!\!\!\prod\limits_{k\in\mathcal{N}_{n}^{(m)}}  \dfrac{f(\mathbf{z}^{(m)}_{n,k}|\mathbf{u}_{n},{\alpha}_{n},\bm{\omega}_{n}^{(m)},\breve{{\mathbf{v}}}_{n}^{(m)})}{f_{\rm false}(\mathbf{z}^{(m)}_{n,k})}.
	\end{multline} 
	As $\prod_{i=1}^{|\mathcal{M}_{n}^{(m)}|}f_{\rm false}(\mathbf{z}^{(m)}_{n,i})$ is a constant on condition that $ {c}_{n}^{(m)}$ is known, we obtain 
	\begin{multline}\label{mmmp2}
	 f(\mathbf{z}^{(m)}_{n}|\mathbf{u}_{n},{\alpha}_{n},\bm{\omega}_{n}^{(m)},{\mathbf{v}}_{n}^{(m)},\mathbf{a}_{n}^{(m)}, {c}_{n}^{(m)}) \propto \!\!\! \prod\limits_{j\in\mathcal{D}_{n}^{(m)}} \!\! \dfrac{f(\mathbf{z}^{(m)}_{n,{a}_{n,j}^{(m)}}|\mathbf{u}_{n},{\alpha}_{n},\bm{\omega}_{n}^{(m)},\tilde{{\mathbf{v}}}_{n}^{(m)})}{f_{\rm false}(\mathbf{z}^{(m)}_{n,{a}_{n,j}^{(m)}})} \\ \times\prod\limits_{k\in\mathcal{N}_{n}^{(m)}}  \dfrac{f(\mathbf{z}^{(m)}_{n,k}|\mathbf{u}_{n},{\alpha}_{n},\bm{\omega}_{n}^{(m)},\breve{{\mathbf{v}}}_{n}^{(m)})}{f_{\rm false}(\mathbf{z}^{(m)}_{n,k})}.
	\end{multline} 
	Eq. \eqref{mmmp2} means that the measurement evaluation is determined in the perspective of legacy and new features on the condition that the total number of measurements is known.

	\subsubsection{Data Association}\label{da}
	
	The joint prior pdf of the data association vectors $\mathbf{a}_{n}^{(m)}$ and $\mathbf{b}_{n}^{(m)}$,  number-of-measurements $ {c}_{n}^{(m)}$, and  state of new features $\breve{{\mathbf{v}}}_{n}^{(m)}$ conditioned on the state of legacy features $\tilde{{\mathbf{v}}}_{n}^{(m)}$ and agent $\mathbf{u}_{n}$, and unknown measurement biases ${\alpha}_{n}$ and $\bm{\omega}_{n}^{(m)}$ is  
	\begin{multline}\label{da1}
 f(\mathbf{a}_{n}^{(m)},\mathbf{b}_{n}^{(m)}, {c}_{n}^{(m)},\breve{{\mathbf{v}}}_{n}^{(m)}|\tilde{{\mathbf{v}}}_{n}^{(m)},\mathbf{u}_{n},{\alpha}_{n},\bm{\omega}_{n}^{(m)})  
	\propto   \Psi(\mathbf{a}_{n}^{(m)},\mathbf{b}_{n}^{(m)})\mu_{\rm new}^{(m)|\mathcal{N}_{n}^{(m)}|}\mu_{\rm false}^{(m)(-|\mathcal{N}_{n}^{(m)}|-|\mathcal{D}_{n}^{(m)}|)}\\
	\times  \!\!\!\!\prod\limits_{j\in\mathcal{D}_{n}^{(m)}}\!\!\!\!\!P_{\rm d}(\mathbf{u}_{n},\mathbf{{p}}_{n,{a}_{n,j}^{(m)}}^{(m\!)})\! \!\!\!\!\prod\limits_{j'\in\bar{\mathcal{D}}_{n}^{(m)}}\!\!\!\!\!\!\left(\!1\!-\!P_{\rm d}(\mathbf{u}_{n},\mathbf{{p}}_{n,j'}^{(m)})\right)\!\!\!\!\!\!
 \prod\limits_{k\in\mathcal{N}_{n}^{(m)}}\!\! \!\!\!\!f_{\rm new}(\breve{{\mathbf{v}}}_{n,k}^{(m)}|\mathbf{u}_{n},\!{\alpha}_{n},\!\bm{\omega}_{n}^{(m)})\!\!\!\!\!\!\prod\limits_{k'\!\in\bar{\mathcal{N}}_{n}^{(m)}}\!\!\!\!\!\! f_D(\breve{{\mathbf{v}}}^{(m)}_{n,k'}\!),
	\end{multline}
	where $\bar{\mathcal{D}}_{n}^{(m)}= \mathcal{K}_{n-1}^{(m)} \backslash {\mathcal{D}}_{n}^{(m)}$, $\bar{\mathcal{N}}_{n}^{(m)}= \mathcal{M}_{n}^{(m)} \backslash {\mathcal{N}}_{n}^{(m)}$, ``$\backslash$" represents the complement operator, $P_{\rm d}(\cdot) \!\!\in\!\! (0,1]$ is 
	the probability that a feature is ``detected" in the sense that it generates a measurement $\mathbf{z}_{n,l}^{(m)}$, and $f_{\rm new}(\cdot)$ represents some pdf of the newly detected features.
	The detailed derivation is given in Appendix \ref{A}. 
	Note that the factorization given in \eqref{da1} is also in the perspective of legacy and new features.

	\subsubsection{Data Fusion}
	The joint posterior distribution of the agent state,  measurement biases, feature state, and data association vectors conditioned on measurements for all $N$ time slots is $f(\mathbf{u}_{1:N},\bm{\alpha}_{1:N},\bm{\omega}_{1:N}, {\mathbf{v}}_{1:N},\mathbf{a}_{1:N},\mathbf{b}_{1:N}|\mathbf{z}_{1:N})$.
	According to \eqref{jointP1}, the joint posterior distribution is the product of \eqref{mark}, \eqref{mmmp2}, and \eqref{da1}.
	As the factorizations of \eqref{mmmp2} and \eqref{da1} 
	are in the perspective of legacy and new features, we rewrite equations \eqref{mmmp2} and \eqref{da1} in a more concise form.
	Note that ${\mathcal{D}}_{n}^{(m)}$ is the set of legacy feature indexes that generate measurement at time $n$. 
	It is implied that for $j \in {\mathcal{D}}_{n}^{(m)}$
	we have $\tilde{r}_{n,j} = 1$ and ${a}_{n,j}^{(m)}\neq 0$. 
	On the contrary,
	$\bar{\mathcal{D}}_{n}^{(m)}$ is the set of legacy feature indexes which do not generate measurement at time $n$, and for $j \in \bar{\mathcal{D}}_{n}^{(m)}$
	we have $\tilde{r}_{n,j} = 1$ and ${a}_{n,j}^{(m)}= 0$.
	We define a function $g(\mathbf{u}_{n},{\alpha}_{n},\bm{\omega}_{n}^{(m)},\tilde{\mathbf{v}}_{n,j}^{(m)},\mathbf{a}^{(m)}_{n,j};\mathbf{z}^{(m)}_{n,j})$, when $\tilde{r}_{n,j} = 1$, and we have 
	\begin{equation}\label{g1}	\!\!g(\mathbf{u}_{n},{\alpha}_{n},\bm{\omega}_{n}^{(m)},\tilde{\mathbf{v}}_{n,j}^{(m)},\mathbf{a}^{(m)}_{n,j};\mathbf{z}^{(m)}_{n,j})=
	\!\!\left\{
	\begin{array}{ll}
	\!\!\dfrac{f(\mathbf{z}^{(m)}_{n,{a}_{n,j}^{(m)}}|\mathbf{u}_{n},{\alpha}_{n},\bm{\omega}_{n}^{(m)},\tilde{{\mathbf{v}}}_{n}^{(m)})P_{\rm d}(\mathbf{u}_{n},\mathbf{{p}}_{n,{a}_{n,j}^{(m)}}^{(m)})}{\mu_{\rm false}^{(m)}f_{\rm false}(\mathbf{z}^{(m)}_{n,{a}_{n,j}^{(m)}})},  \!\!& \!  {a}_{n,j}^{(m)}\!\neq\! 0,\\
	\!\!1-P_{\rm d}(\mathbf{u}_{n},\mathbf{{p}}_{n,j}^{(m)}), \!\!&\! {a}_{n,j}^{(m)}\!=\! 0,
	\end{array} \right.
	\end{equation}
	and when $\tilde{r}_{n,j}\!\! =\!\! 0$, we have
	$
	g(\!\mathbf{u}_{n},{\alpha}_{n},\bm{\omega}_{n}^{(m)}\!,\!\tilde{\mathbf{v}}_{n,j}^{(m)}\!,\!\mathbf{a}^{(m)}_{n,j};\mathbf{z}^{(m)}_{n,j}\!)\!\!=\!\!1.
	$
	Moreover, ${\mathcal{N}}_{n}^{(m)}$ denotes the set of measurement indexes generated by new features, which means that for $k \in {\mathcal{N}}_{n}^{(m)}$ we have $\breve{r}_{n,k}\!\!=\!\!1$ and ${b}_{n,k}^{(m)}\!\!=\!\!0$. By contrast,
	$\bar{\mathcal{N}}_{n}^{(m)}$ denotes the set of measurements that are not generated by new features, and we have $\breve{r}_{n,k}\!\!=\!\!0$, for $k \in \bar{\mathcal{N}}_{n}^{(m)}$.
	We define a function $h(\mathbf{u}_{n},{\alpha}_{n},\bm{\omega}_{n}^{(m)}\!\!,\!\breve{\mathbf{v}}_{n,k},\mathbf{b}^{(m)}_{n,k};\mathbf{z}^{(m)}_{n,k})$, when $\breve{r}_{n,k}=1$, we have 
	\begin{equation}\label{h1}
	\!\!\!h(\mathbf{u}_{n},{\alpha}_{n},\bm{\omega}_{n}^{(m)}\!\!,\!\breve{\mathbf{v}}_{n,k},\mathbf{b}^{(m)}_{n,k};\mathbf{z}^{(m)}_{n,k})=
	\!\!\left\{
	\begin{array}{ll}
	\!\!\!  0,  &  \!\!   {b}_{n,k}^{(m)}\!\!\neq\!\! 0,\\
	\!\!\!   \dfrac{\mu_{\rm new}^{(m)}\!f_{\rm new}\!(\breve{\mathbf{{v}}}_{n,k}^{(m)}\!| \mathbf{u}_{n},\!{\alpha}_{n},\!\bm{\omega}_{n}^{(m)}\!)\!f(\mathbf{z}^{(m)}_{n,k}\!|\mathbf{u}_{n}\!,\!{\alpha}_{n}\!,\!\bm{\omega}_{n}^{(m)}\!,\!\tilde{\mathbf{v}}_{n}^{(m)}\!)}{\mu_{\rm false}^{(m)}f_{\rm false}(\mathbf{z}^{(m)}_{n,k})}, &  \!\!    {b}_{n,k}^{(m)}\!\!=\!\! 0,
	\end{array} \right.
	\end{equation}
	and  when $\breve{r}_{n,k}\!\!=\!0$, we have  
	$
	h(\mathbf{u}_{n},{\alpha}_{n},\bm{\omega}_{n}^{(m)}\!\!,\!\breve{\mathbf{v}}_{n,k},\mathbf{b}^{(m)}_{n,k};\mathbf{z}^{(m)}_{n,k})\! =\! f_D(\breve{\mathbf{{v}}}_{n,k}).
	$
	Finally, the joint posterior pdf is given by{\begingroup\makeatletter\def\f@size{12}\check@mathfonts
			\def\maketag@@@#1{\hbox{\m@th\normalsize\normalfont#1}}\setlength{\arraycolsep}{0.0em}\setlength{\arraycolsep}{0.0em}
			\begin{eqnarray}\label{final}\hspace{-0.7cm}
			\begin{array}{ll}
		& f(\mathbf{u}_{1:N},\bm{\alpha}_{1:N},\bm{\omega}_{1:N},{\mathbf{v}}_{1:N},\mathbf{a}_{1:N},\mathbf{b}_{1:N}|\mathbf{z}_{1:N})\\
		 \propto &  \underbrace{\prod\limits_{n=1}^{N}
				\!\! f(\mathbf{u}_{n}|\mathbf{u}_{n\!-\!1}) f(\alpha_{n}|\alpha_{n\!-\!1}) \!\!\prod\limits_{m=1}^{M} \!\! f(\omega_{n}^{(m)}|\omega_{n\!-\!1}^{(m)})\!\!\prod\limits_{j=1}^{|\mathcal{K}_{n-1}^{(m)}|} \!\! f({\tilde{\mathbf{v}}}^{(m)}_{n,l}|{\mathbf{v}}^{(m)}_{n\!-\!1,l})}_{(a)}\\
			&\times \underbrace{\prod\limits_{n=1}^{N}\prod\limits_{m=1}^{M} 
				\Psi(\mathbf{a}_{n}^{(m)}\!,\!\mathbf{b}_{n}^{(m)}\!)}_{(c)}  \underbrace{\!\!\prod\limits_{j=1}^{|\mathcal{K}_{n-1}^{(m)}|} \!\! g(\mathbf{u}_{n},\!{\alpha}_{n},\!\bm{\omega}_{n}^{(m)}\!,\!\tilde{\mathbf{v}}_{n,j}^{(m)}\!,\!\mathbf{a}^{(m)}_{n,j}\!;\!\mathbf{z}^{(m)}_{n,j})\! \!\prod\limits_{k=1}^{|\mathcal{M}_{n}^{(m)}|}\!\!h(\mathbf{u}_{n},\!{\alpha}_{n},\!\bm{\omega}_{n}^{(m)}\!,\!\breve{\mathbf{v}}^{(m)}_{n,k}\!,\!\mathbf{b}^{(m)}_{n,k}\!;\!\mathbf{z}^{(m)}_{n,k})}_{(b)},
			\end{array}
			\end{eqnarray}\setlength{\arraycolsep}{5pt}\endgroup}where (a), (b), and (c) corresponds to those in \eqref{jointP1}.

	\begin{remark}
		The AOA, TOA, and RSS measurements can be combined according to the agent's hardware facilities and measurement acquisition conditions.
		The measurements can be easily added to or removed from the
		BP-based SLAM algorithm according to \eqref{final}.
		For example, if TOA measurements are unavailable, then we obtain $f(\omega_{1}^{(m)})=1$ and $f(\omega_{n}^{(m)}|\omega_{n-1}^{(m)})=1$, which is similar for AOA and RSS measurements. \vspace{-0.2cm}
	\end{remark}
	
	\begin{figure}
		\centering
		\includegraphics[scale=0.75]{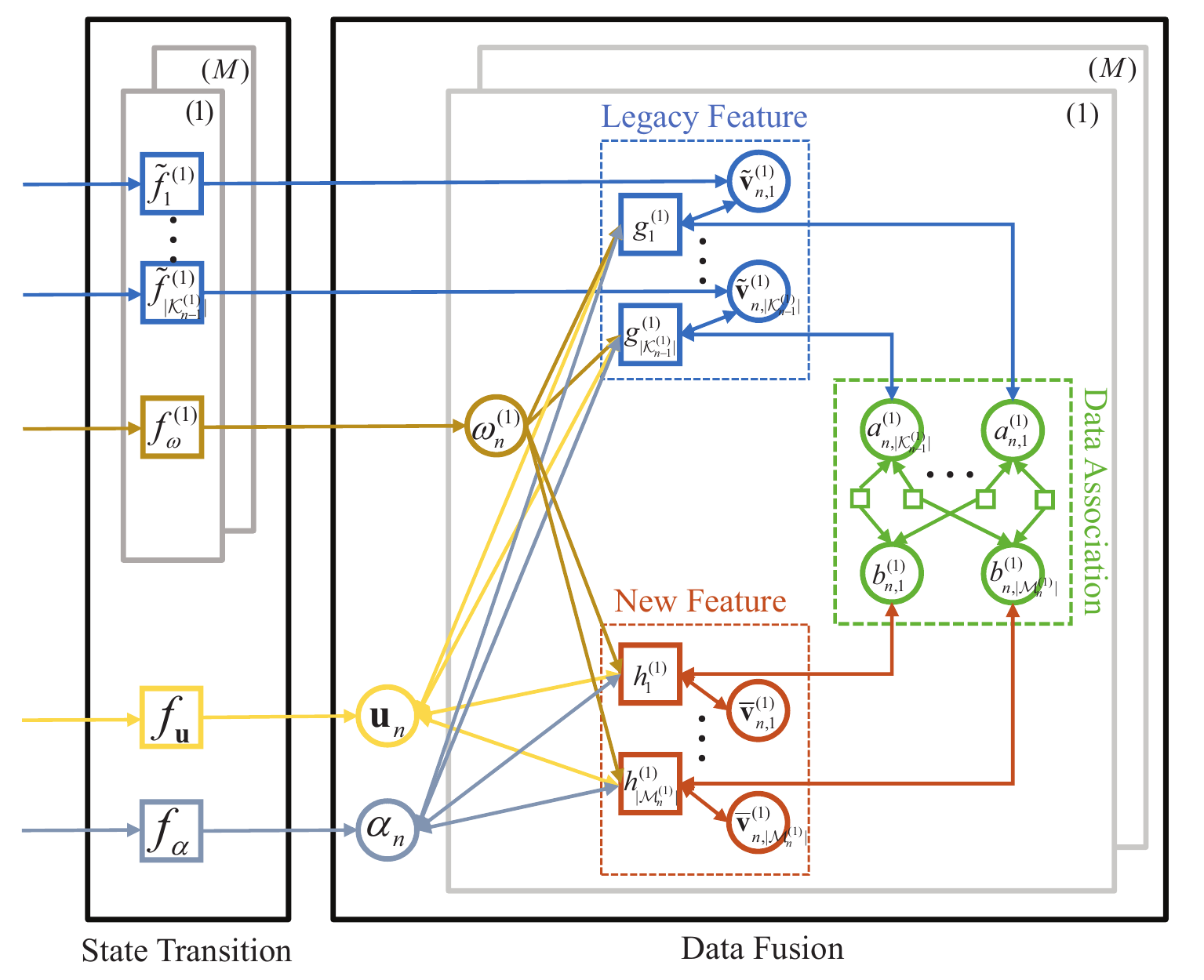}
		\caption{Factor graph represents the factorization in \eqref{final}.} \label{fig:fg}
		\vspace{-0.4cm}
	\end{figure}
	
	The factor graph representing the factorization in \eqref{final} is depicted in Fig. \ref{fig:fg}.
	The messages are only sent forward in time.
	We adopt a widely accepted order \cite{BP1} that the messages should be computed.
	The messages first undergo the state transition phase, in which the messages from time $n-1$ pass through factor nodes $f_{\mathbf{u}}$, $f_{{\alpha}}$, $f_{{\omega}}^{(m)}$, and $\tilde{f}^{(m)}_{l}$ for $m=1,\ldots,M$, and $l = 1,\ldots, |\mathcal{K}_{n-1}^{(m)}|$ to generate 
	prediction messages.
	Then, measurement evaluation calculations are processed by factor node $g^{(m)}_{l}$ for legacy features and factor node $h^{(m)}_{l'}$ for new features in parallel, where $l' = 1,\ldots, |\mathcal{M}_{n}^{(m)}|$.
	Next, the output messages of factor nodes $g^{(m)}_{l}$ and $h^{(m)}_{l'}$ are passed to the data association variable nodes $a_{n,l}^{(m)}$ and $b_{n,l'}^{(m)}$, respectively. 
	The messages are calculated iteratively among $a_{n,l}^{(m)}$ and $b_{n,l'}^{(m)}$, and this process is called the loopy data association phase.
	After the last iteration, the messages are passed back from $a_{n,l}^{(m)}$ to $g^{(m)}_{l}$ and from $b_{n,l'}^{(m)}$ to $h^{(m)}_{l'}$. 
	The messages are subsequently updated by factor nodes $g^{(m)}_{l}$ and $h^{(m)}_{l'}$. 
	Finally, the messages are fused at variable nodes $\mathbf{u}_n$, $\omega^{(m)}_n$, $\alpha_n$, $\tilde{\mathbf{{v}}}_{n,l'}^{(m)}$, and $\breve{\mathbf{{v}}}_{n,l'}^{(m)}$.
	Once the messages are available, the belief approximating the desired marginal posterior pdfs are obtained.
	
	\subsection{Probabilistic Crowdsourcing Mechanism}\label{cs}
	
	In this subsection, we aim at further improving the performance of the proposed plug-and-play SLAM by the cooperation among agents. 
	We establish a crowdsourcing mechanism to construct and refine the radio map.
	Moreover, agents can download the generated radio map from the cloud to assist and accelerate their own communication and SLAM processes.
	In the proposed crowdsourcing mechanism, we assume that the mobile agents would honestly provide the information they obtain. A reputation and pseudonym manager can be introduced in the proposed system for future research, according to \cite{M1} and \cite{M2}, to achieve trust evaluation and privacy preservation.

	\begin{Def}
		We define a set $\mathcal{L}_k$, which contains the features' state estimated by agent $k$, as the \textbf{local radio feature map (LRF-Map)} of agent $k$, given as 
		\begin{equation}
		\mathcal{L}_k = 
		\Big\{ \mathcal{L}_k^{(1)}, \ldots, \mathcal{L}_k^{(M)} \Big\},  
		\end{equation}
		and
		\begin{equation}
		\mathcal{L}_k^{(m)} = \Big\{  [\tilde{\mathbf{p}}^{(m)}_{k,n',l},
		{P_{\rm e}}_{k,n',l}^{(m)}
		]\Big|  l=1,\ldots, |\mathcal{K}_{k,n'}^{(m)}| \Big\}, 
		\end{equation}
		where $n'$ denotes the convergent time of the SLAM algorithm at agent $k$, and ${P_{\rm e}}_{k,n',l}^{(m)}$ represents the existence probability of feature $\tilde{\mathbf{p}}^{(m)}_{k,n',l}$ calculated by \eqref{marginal}.
		Note that we add the subscript $k$ in $\tilde{\mathbf{p}}^{(m)}_{k,n',l}$ and $\mathcal{K}_{k,n'}^{(m)}$ to distinguish different agents.
		Moreover, 
		we define a set $\mathcal{O}$ in the cloud as the \textbf{open radio feature map (ORF-Map)}, which contains radio features in the ROI.
		The ORF-Map $\mathcal{O}$ is a weighted combination of the LRF-Map $\mathcal{L}_k$ for $k=1,\ldots,K$, where $K$ denotes the total number of accessed agents.	
		The ORF-Map is a dynamic set and keeps receiving information from agents for updates.
	\end{Def}
	\begin{figure}
		\centering
		\includegraphics[scale=0.5]{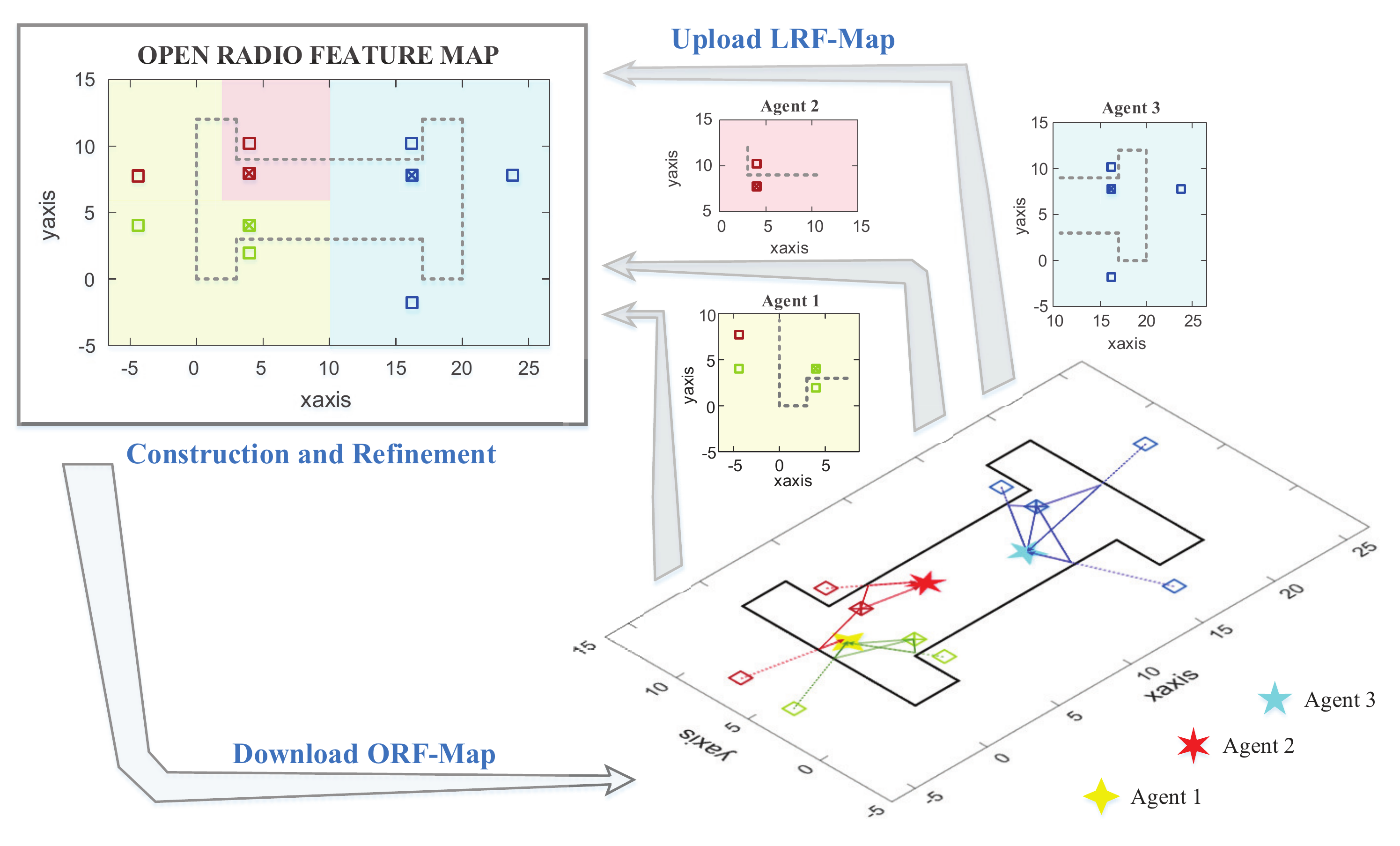}
		\caption{The proposed crowdsourcing method which consists of three parts, including the construction, usage, and refinement of the open radio feature map.} \label{fig:cs}
	\end{figure}
	
	\begin{figure}
		\centering
		\vspace{-0.2cm}
		\includegraphics[scale=0.68]{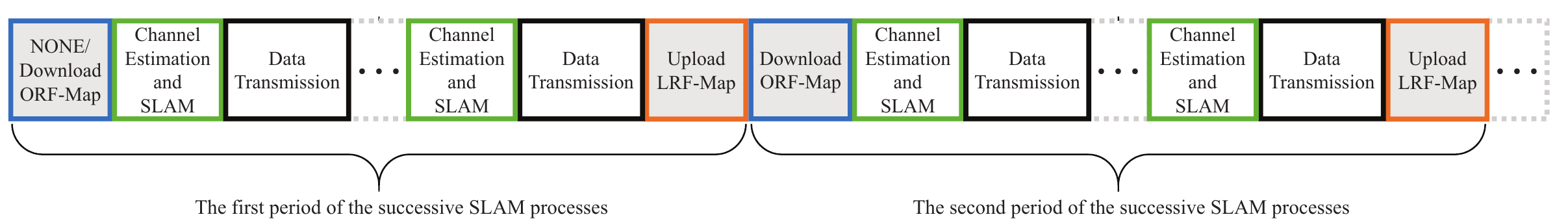}
		\caption{Physical layer frame structure of an agent corresponding to the proposed crowdsourcing mechanism.} \label{fig:frame}
		\vspace{-0.2cm}
	\end{figure}
	The idea of the proposed crowdsourcing mechanism is shown in Figs. \ref{fig:cs} and \ref{fig:frame}.
	At the very beginning, the ORF-Map $\mathcal{O}$ is empty.
	The frame structure of each agent is illustrated in Fig. \ref{fig:frame}, where no prior information can be download at the very beginning.
	Channel estimation and SLAM are executed at the same time slots on the agent side.
	The initial phase involves constructing an ORF-Map of a new environment.
	Each agent can analyze the environment separately through its own received RF signals and sensors by using the method proposed in Section \ref{mpp}. The agents are not necessary to enter the environment at the same time.
	The data transmission comes after each channel estimation and SLAM slot.
	After several time slots of channel estimation and SLAM processes, 
	a threshold is set to ensure that the SLAM result of each agent converges.
	The radio environment observed by each agent from their own perspective is incomplete or biased.
	The LRF-Maps obtained by three agents are subsets of the total radio features in the ROI (Fig. \ref{fig:cs}).
	According to the frame structure in Fig. \ref{fig:frame}, each agent uploads its LRF-Map to the cloud when the mapping result converges.
	As illustrated in Fig. \ref{fig:cs}, the fusion of three LRF-Maps can generate a complete ORF-Map in the cloud.
	Let $K_1$ represent the number of the first batch of accessed agents in the initial construction phase.
	Then, we construct the ORF-Map $\mathcal{O}$ at the cloud by
	\begin{equation}\label{union}
	\mathcal{O}^{(m)} = \bigcup\limits_{k=1}^{K_1} \mathcal{{L}}_k^{(m)}, 
	\end{equation}
	for $m=1,\ldots,M$, and 
	\begin{equation}\label{uunion}
	\mathcal{O} = 
	\Big\{ \mathcal{O}^{(1)}, \ldots, \mathcal{O}^{(M)} \Big\}.  
	\end{equation}
	The generated ORF-Map is stored at the cloud.
	
	The established ORF-Map can be downloaded with various applications. 
	Here, we take SLAM as an example.
	For newly accessed agents, 
	the ORF-Map is downloaded at the first slot according to the frame structure in Fig. \ref{fig:frame},
	and the downloaded information can be considered as a good initial value.
	For the already accessed agents, the downloaded information can be used to complete their own LRF-Maps.
	Specifically, the ORF-Map can provide candidate legacy features to each agent in need.
	Let $n$ denote the current time slot,
	for agent $k$, we have $\mathcal{K}_{k,n\!-\!1}^{(m)} = \mathcal{O}^{(m)}$ for $m=1, \ldots, M$.
	Note that the agent only needs to download features corresponding to the anchors it can access.
	The data association of measurements obtained by agent $k$ at the current time slot $n$ and legacy features in $\mathcal{K}_{k,n\!-\!1}^{(m)}$ is denoted by two data association vectors given in \eqref{DA1} and \eqref{DA2} and executed at the agent side. 
	Then, the SLAM is processed at agent $k$ using the method proposed in Section \ref{mpp}.
	Note that although data association is executed at the agent side, with the accumulation of time, the ORF-Map is constantly updated, thereby solving the double-counting issue in the ORF-Map as discussed in detail in Remark \ref{dc}.

	
	The ORF-Map in the cloud is constantly updated.
	According to the frame structure in Fig. \ref{fig:frame}, the agent keeps uploading the newly estimated LRF-Map to the cloud over time.
	Note that the LRF-Maps obtained by different agents have distinct qualities.
	On the one hand, agents are equipped with different hardware conditions, which affect the quality of extracted measurements.
	On the other hand, agents visiting at different time slots inherit a priori information with different precisions.
	During the refinement phase of the crowdsourcing mechanism,
	we introduce a weight coefficient $\varphi_{k,n'}$ to indicate the reliability of the LRF-Map obtained by the $k$-th agent at the time slot $n'$.
	The weighted LRF-Map generated by the $k$-th agent is calculated by 
	\begin{equation}
	{\mathcal{L}}_k^{(m)} = \Big\{  [\tilde{\mathbf{p}}^{(m)}_{k,n',l},\varphi_{k,n'}{P_{\rm e}}_{k,n',l}^{(m)}]\Big| l=1,\ldots, |\mathcal{K}_{k,n'}^{(m)}| \Big\}, 
	\end{equation}
	where $m=1,\ldots,M$, and the product $\varphi_{k,n'}{P_{\rm e}}_{k,n',l}^{(m)}$ represents the reliability of feature $\tilde{\mathbf{p}}^{(m)}_{k,n',l}$.
	After the weighted LRF-Map estimated by agent $k$ is uploaded to the cloud, it is combined with the existing ORF-Map to generate a complete radio feature map by
	\begin{equation}\label{union2}
	\mathcal{O}^{(m)} = \mathcal{O}^{(m)} \cup \mathcal{L}_{k}^{(m)}. 
	\end{equation}
	The ORF-Map becomes more accurate over time given the increase in its contributions, thereby providing more precise prior information.
	On the basis of the observation that the LRF-Map estimated by the former agents is less accurate than that of the later agents, the weight of the estimated LRF-Map is set to be proportional to the upload time (that is, the convergent time of the estimated LRF-Map).
	For weight coefficients $\varphi_{\ast,n_1}$ and $\varphi_{\ast,n_2}$, where $n_1<n_2$ and $\ast$ denotes any agents, we have $\varphi_{\ast,n_1}<\varphi_{\ast,n_2}$.
	Namely, the weights of the LRF-Map estimated and uploaded later are larger than that of the former. 
	
	As the ORF-Map $\mathcal{O}$ is stored at the cloud,
	to prevent the continuous increase of data in set $\mathcal{O}$, pruning is executed according to the reliability of each feature.
	Specifically,
	let ${P_{\rm w}}_{n,l}$ denote the reliability of the $l$-th feature in $\mathcal{O}^{(m)}$ at time slot $n$.
	Then, the elements in $\mathcal{O}^{(m)}$ can be represented as 
	$[\tilde{\mathbf{p}}^{(m)}_{n,l},{P_{\rm w}}_{n,l}]$ for $l=1,\ldots, |\mathcal{O}^{(m)}|$.
	If ${P_{\rm w}}_{n,l}< P_{\rm{threshold}}$, the $l$-th feature $[\tilde{\mathbf{p}}^{(m)}_{n,l},{P_{\rm w}}_{n,l}]$ is deleted in $\mathcal{O}^{(m)}$.
	The pruning process can save storage resource and exclude unreliable features.
	
	\begin{remark} \label{dc}
		The first challenge of crowdsourcing is the ``double count", which means that some different estimations in the set $\mathcal{O}$ may correspond to the same feature in reality because of the estimation error of different agents.	
		Although the union operation in \eqref{union} leads to the double count problem, the proposed probabilistic data association-based crowdsourcing mechanism can solve this problem.
		Specifically, as we have $\mathcal{K}_{k,n\!-\!1}^{(m)} = \mathcal{O}^{(m)}$ for $m=1, \ldots, M$, 
		the double count problem happens in the set of the legacy features for each agent.
		With the probability data association mentioned in Section \ref{mpp}, each agent can associate its measurements to the more reliable legacy features and thus can 
		filter out legacy features with large errors.
		Finally, the number of estimated features by each agent converges to the number of measurements approximately.
		Moreover, the feature estimations become increasingly accurate over time, where accurate estimations of the same feature from different agents converge to the same location.
	\end{remark}

	\begin{remark}
		The second challenge of crowdsourcing is ``data reliability", which indicates that valuable and unreliable data are mixed together during crowdsourcing.
		The proposed probabilistic crowdsourcing mechanism can solve this problem by calculating the reliability (weighted existence probability) of each feature.
		Through the pruning process, the unreliable data (features with small weighted existence probability) are filtered out.	
	\end{remark}

	\begin{remark}
		The proposed plug-and-play mechanism for SLAM with crowdsourcing has the following advantages, which make it applicable for IoT devices.	 
		First, although the IoT systems exhibits heterogeneity and different devices have different capabilities to obtain measurements,
		the proposed measurement plug-and-play mechanism can work flexibly with different categories of measurements,
		even for the partial category (such as one of AOA, TOA, or RSS) of measurements.
		Second, by abstracting the radio environment into the radio feature map, the information exchange format is unified to the features' 2-D coordinates and existence probability, thereby reducing the amount of data and solving the challenge of information exchange among IoT devices. 
		Third, as the construction, refinement, and storage of the ORF-Map are carried out at the cloud, the computing and storage resource requirements of each agent can be relaxed.
		Moreover, as only the LRF-Map is uploaded in the proposed mechanisms and the trajectory is not required, 
		the privacy of each agent can be protected.
	\end{remark}


	\section{Numerical Result}\label{s6}
	\subsection{Simulation Setup}
	In this section, we assess the performance of the proposed algorithm.
	We use the floor plan shown in Fig. \ref{fig:floorplan}, the size of which is approximately $20\times 12\  \rm{m}^2$.
	The ROI is a circular disk with a radius of $40$ m.
	Most of the messages leaving the factor nodes cannot be solved in a closed form owing to the contained integrals. 
	We use particle-based implementation to approximate the continuous messages \cite{BP1}.
	
	\subsubsection{Measurement Model}
	True AOA, TOA, and RSS are generated according to \eqref{AOA}, \eqref{TOA}, and \eqref{RSS}, respectively. 
	The measurement noise follows a zero-mean additive white Gaussian distribution, and the standard deviation is set to $\sigma_a=1$ degree for the AOA, $\sigma_t=0.15$ m for the TOA, and $\sigma_r=2.5$ dBm for the RSS.
	
	\subsubsection{State Transition Model}
	The state transition of the agent is given by   $\mathbf{u}_{n}^{\rm T} = \mathbf{A} \mathbf{u}^{\rm T}_{n-1}+ \mathbf{d}_{n}$ in Section \ref{st}.
	In the following simulations, we set the variance of the driving process $\mathbf{d}_{n}$ to $\sigma_d^2=0.0278$ and the sampling period to $\Delta T=1$ s.
	For features, the state-transition pdfs is given by Dirac delta functions. 
	According to \cite{BP2}, we introduced a small driving process for the sake of numerical stability.
	Here, we have
	$\tilde{{\mathbf{v}}}^{(m)}_{n,l}={\mathbf{v}}^{(m)}_{n-1,l}+\boldsymbol{\varpi}^{(m)}_{n,l}$,
	and $\boldsymbol{\varpi}^{(m)}_{n,l}$ follows the independently identically Gaussian distribution across $n$ with zero-mean and covariance matrix $\sigma_\varpi^2 \mathbf{I}$, and we set $\sigma_\varpi^2 = 10 ^ {-8}$.

	\subsubsection{Initialization}
	The initial location of the agent is not known exactly, as we know only the approximate location of the entrance.
	The particles for the initial agent location are drawn from a uniform distribution within a circle, with $[x_{{e}},y_{{e}}]$ as the center and $\lambda$ as the radius, where $[x_{{e}},y_{{e}}]$ is the center of the actual entrance.
	In the following simulations, $\lambda$ is set to $0.5$ m. 
	The particles for the initial value of the measurement biases, such as the orientation bias, clock bias, and RSS model parameters, are drawn from a uniform distribution with ranges of $\alpha_n\! \sim\!  \mathcal{U}[-0.5, 0.5]\ {\rm rad}$, $\omega_n^{(m)} \!\sim\!  \mathcal{U}[0, 50]\ {\rm m,}$  $\xi_{n,l}^{(m)}\sim \mathcal{U}[-45, -25]\ {\rm dBm}$, and $\beta_{n,l}^{(m)}\sim \mathcal{U}[2, 5]$.
	The parameters involved in the algorithm are given as  
	$P_{\rm d}=0.95$, $P_{\rm s}=0.999$, $\mu_{\rm false}=1$, $\mu_{\rm new}=10^{-4}$; the unreliability threshold is set to $ 10^{-4}$; the detection threshold is set to $0.5$; and the number of particles is set to $10^5$. \footnote{
	When $N_s=10^5$, and AOA and TOA measurements are used,
		the average runtime per time step for the proposed SLAM method is 
		$1.15$ seconds for a MATLAB implementation on a desktop computer with a 3.3 GHz Intel(R) Xeon(R) W-2155 CPU and 64 GB of RAM, using Windows 10 and MATLAB 2018b (64-bit).}
	
	\begin{figure}[t]
	\vspace{-0.2cm}
	\centering
	\begin{minipage}[t]{0.48\textwidth}
		\hspace{-0.32cm}
		\includegraphics[scale=0.36]{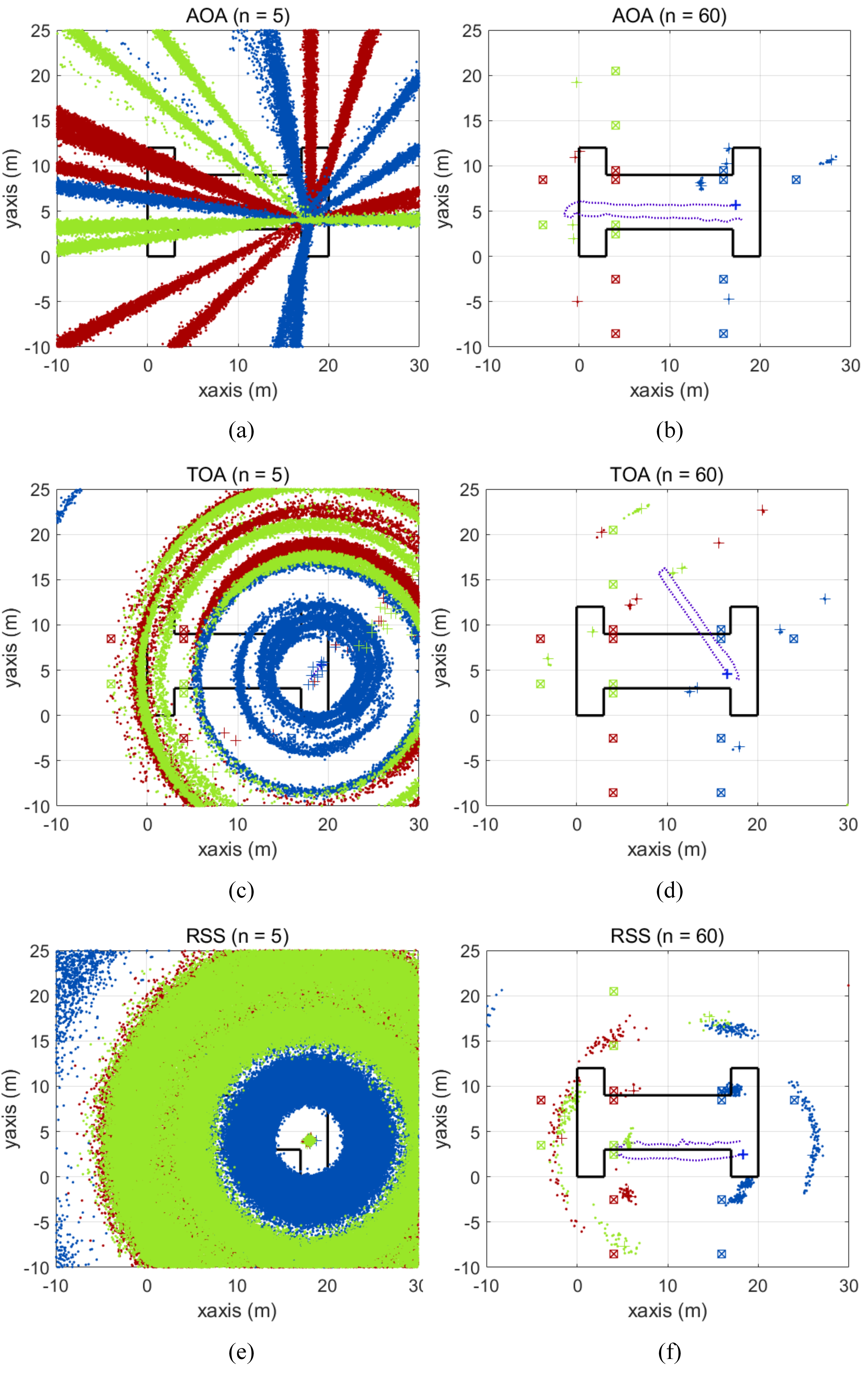}
		\caption{SLAM results for an agent with single category of measurements without measurement biases.}\label{fig:wo}
	\end{minipage}
	\hspace{0.3cm}
	\begin{minipage}[t]{0.48\textwidth}
		\hspace{-0.32cm}
		\includegraphics[scale=0.363]{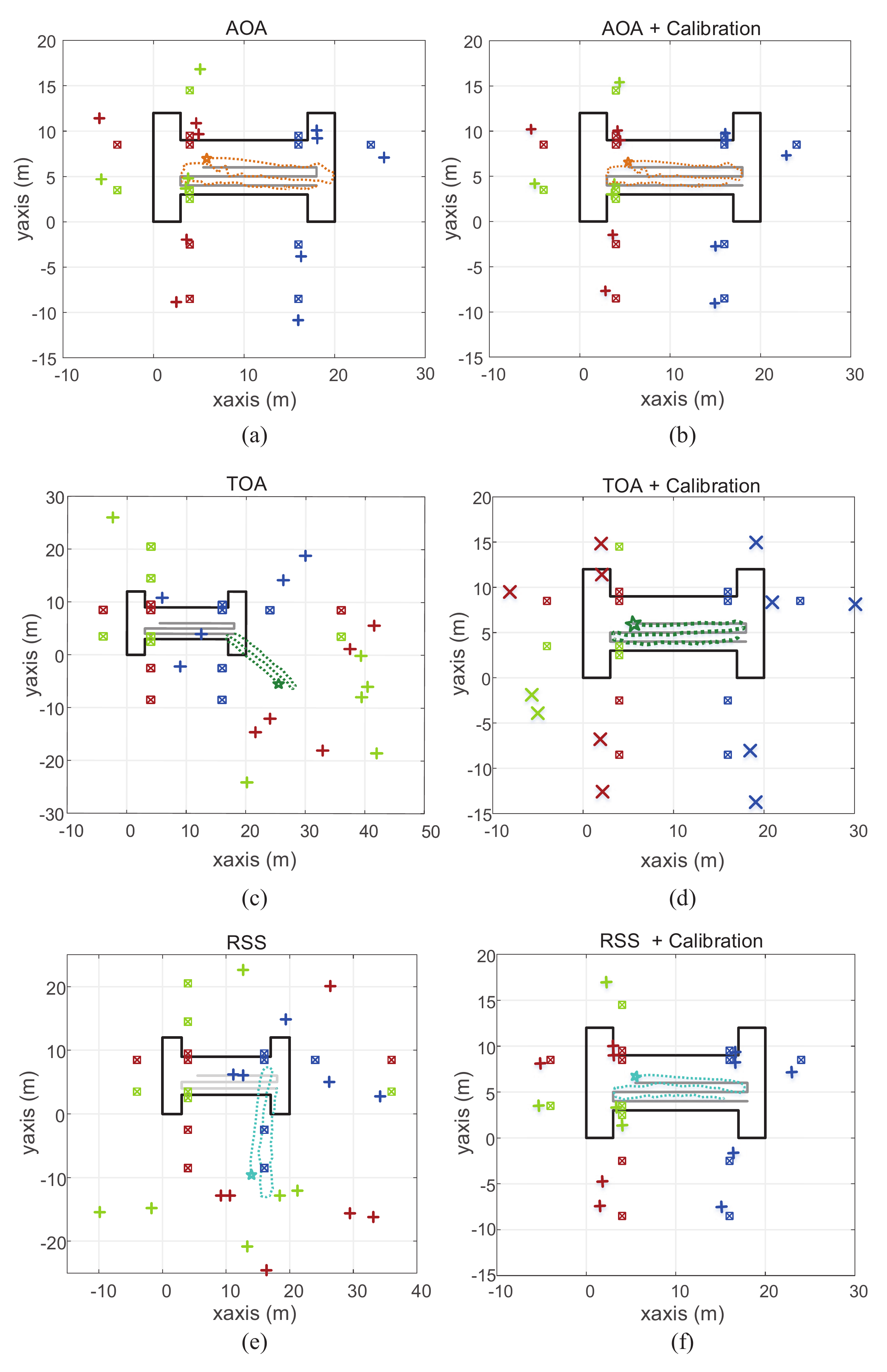}
		\caption{SLAM results for an agent with single category of measurements.}\label{fig:s}
	\end{minipage}   
\end{figure}
	
	\subsection{Performance of the Measurement Plug-and-Play Mechanism}

	\subsubsection{Without Measurement Biases}
	Fig. \ref{fig:wo} shows the SLAM performance of the AOA, TOA, and RSS measurements, for which we assume that the orientation bias, clock bias, and RSS model parameters are perfectly known.
	This experiment seeks to analyze the characteristics of different measurements, such as the AOA, TOA, and RSS measurements.
	The particles generated by AOA measurements are radially distributed (Fig. \ref{fig:wo} [a]).
	The particles generated by the TOA and RSS measurements are distributed in concentric circles and rings (Figs. \ref{fig:wo} [c] and [e]), respectively.
	At $n=60$, most of the features converge for the AOA and TOA measurements (Figs. \ref{fig:wo} [b] and [d]).
	Compared with the AOA and TOA measurements, RSS has a slower convergence speed because of more unknowns (Fig. \ref{fig:wo} [f]).
	As no prior information of the floor plan, anchors, or agents are available, a single measurement category is unable to achieve accurate SLAM results.
	The trend of the overall rotation and deviation of the SLAM result by a single measurement category is explained in Section \ref{w}.

	\subsubsection{With Measurement Biases}\label{w}
		\begin{figure}
		\centering
		\vspace{-0.2cm}
		\includegraphics[scale=0.45]{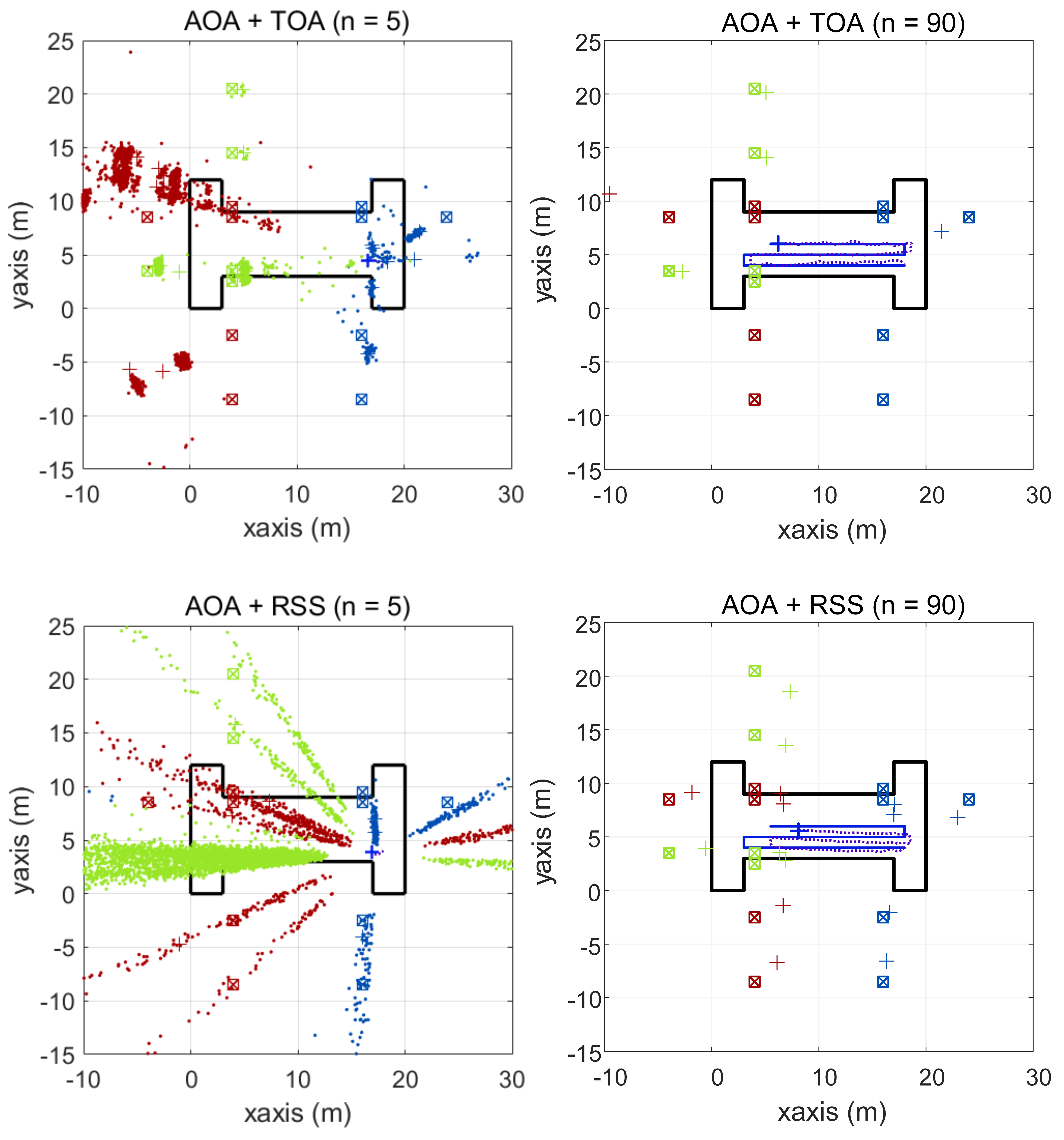}
		\caption{SLAM results for an agent with mixed categories of measurements.}\label{fig:w1}
	\end{figure}
	Fig. \ref{fig:s} shows the SLAM results for the AOA, TOA, and RSS measurements, for which we assume that the orientation bias, clock bias, and RSS model parameters are unknown.
	The SLAM results indicate a trend of overall scaling (Fig. \ref{fig:s} [a]),  rotation (Fig. \ref{fig:s} [c]), and scaling and rotation (Fig. \ref{fig:s} [e]), 
	when only the AOA, TOA, and RSS measurements are used, respectively.
	The calibrated SLAM results, with the calibration of the built-in sensors, such as the gyro (rotation calibration) and accelerator (scaling calibration), are shown in Figs. \ref{fig:s} (b), (d), and (f).
	The positioning performance is satisfactory, as shown in Figs. \ref{fig:s} (b), (d), and (f). However, the mapping error is large owing to the measurement biases.

	Fig. \ref{fig:w1} presents the SLAM results for mixed measurements (AOA + TOA and AOA + RSS) with unknown measurement biases and without sensor calibration.
	When $n=5$, AOA + TOA has a faster convergence speed than AOA + RSS.
	For mapping, as the mapping deviation between the AOA and TOA measurements is large owing to the unknown clock bias, the outcome typically does not converge to an acceptable mapping result when mixed AOA and TOA measurements are employed, thereby causing feature loss (red and blue features in Fig. \ref{fig:w1} for AOA + TOA when $n=90$). 
	Although a slight scaling is observed for AOA + RSS, feature loss does not occur (Fig. \ref{fig:w1} for AOA + RSS when $n=90$).
	For positioning, AOA + TOA outperforms AOA + RSS, as RSS measurements have substantial unknown parameters.

		\begin{figure}
		\vspace{-0.2cm}
		\centering
		\includegraphics[scale=0.5]{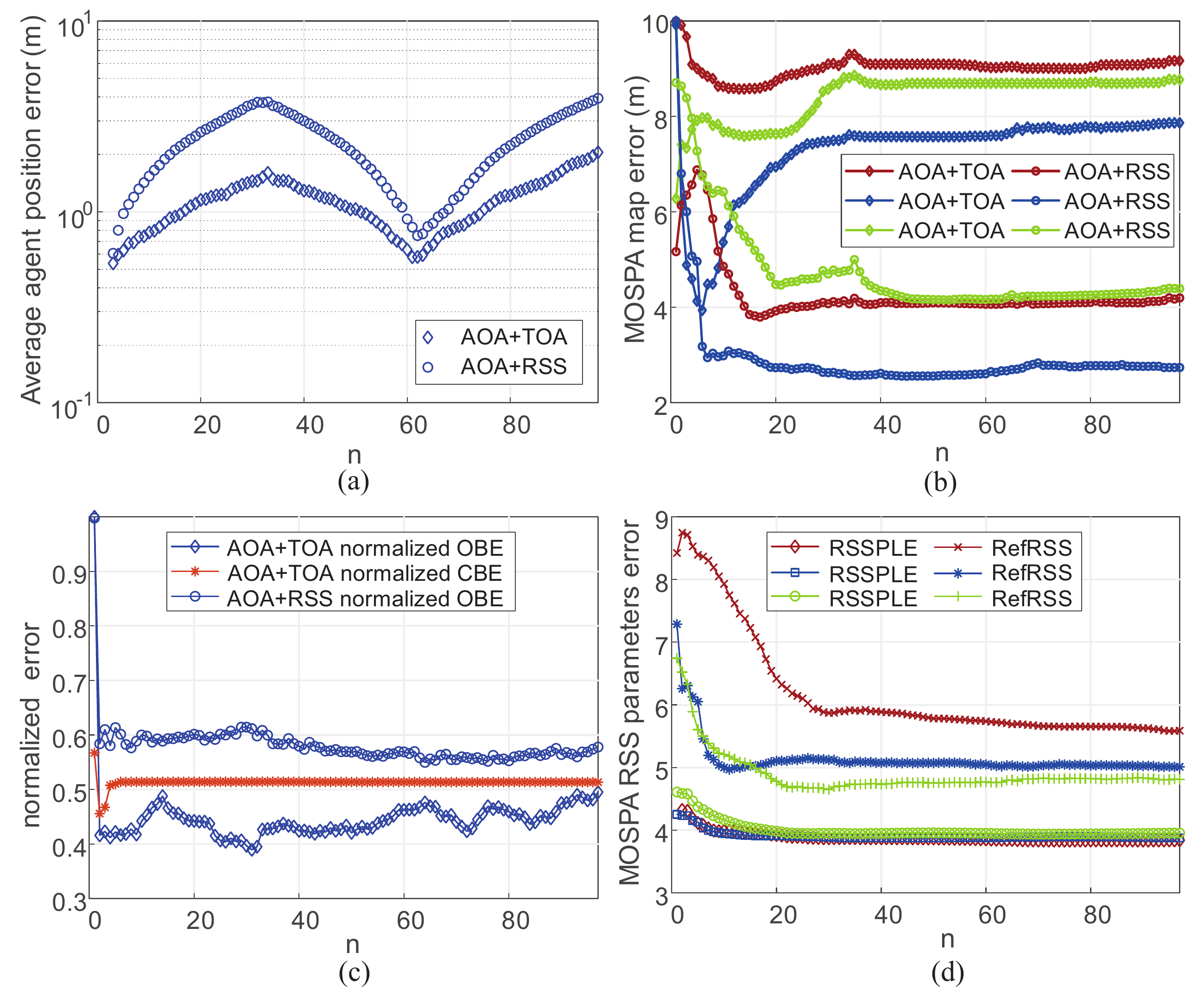}
		\caption{SLAM results obtained from
			$100$ independent Monte Carlo simulations for an agent with mixed categories of measurements.}\label{fig:w2}
	\end{figure}

	The results in Figs. \ref{fig:w2} (a)-(d) are obtained from
	$100$ independent Monte Carlo simulations.
	Let OBE denote the error of the orientation bias estimate and CBE represent the error of the clock bias estimate.
	The Euclidean distance-based mean optimal sub-pattern assignment (MOSPA) metric \cite{BP1} is calculated to evaluate the mapping performance.
	For average positioning error, AOA + TOA outperforms AOA + RSS (Fig. \ref{fig:w2} [a]).
	For average mapping performance, AOA + RSS is more accurate than AOA + TOA (Fig. \ref{fig:w2} [b]).  
	These results are consistent with those shown in Fig. \ref{fig:w1}.  
	The measurement biases are estimated during SLAM (Figs. \ref{fig:w2} [c] and [d]). 
	The orientation and clock biases almost converge at $n = 2$ (Fig. \ref{fig:w2} [c]). 
	The RSS parameters approximately converge at $n = 20$ (Fig. \ref{fig:w2} [d]). 
	This result also explains the phenomenon that TOA and AOA measurements converge faster than RSS measurements.

	Next, we compare the proposed method with the classic BP-SLAM method without  measurement biases, which is denoted as ``Without Bias Est.".
		Table \ref{mmle} compares the maximum value of the mean absolute error (MAE) of the agent trajectory, and
		this metric measures the maximum deviation of the estimated trajectory.
		Table \ref{MOSPA} compares the MOSPA of the physical and virtual anchors,
		and this metric evaluates the mapping performance.
		The numerical results show that the proposed method outperforms the ``Without Bias Est." in both localization and mapping.
		The performance gain increases with the bias level.
		Specifically, when clock and orientation biases are set to $10$ m and $0.5$ rad,
		the proposed method can achieve $68\%$ and $76\%$ performance gain in localization and mapping, respectively.
\begin{table}
	\renewcommand{\arraystretch}{1.5}
	\centering
	\fontsize{10}{10}\selectfont
	\caption{Maximum of the MAE ({\rm m}) of agent trajectory.}\label{mmle}
	\begin{threeparttable}	
		\begin{tabular}{cccc}
			\toprule
			\multirow{2}{*}{Bias Level}& $\omega_n^{(m)}=1$ m  & $\omega_n^{(m)}=5$ m & $\omega_n^{(m)}=10$ m \\
			& $\alpha_n=0.1$ rad & $\alpha_n=0.2$ rad & $\alpha_n=0.5$ rad \\
			\hline
			Proposed Method & {\bf 1.58} & {\bf 1.75} & {\bf 2.31} \\
			Without Bias Est. & 1.98 & 2.93 & 7.25 \\	
			\bottomrule			
		\end{tabular}
	\end{threeparttable}	
\end{table}

\begin{table}
	\renewcommand{\arraystretch}{1.5}
	\centering
	\fontsize{10}{10}\selectfont
	\caption{MOSPA ({\rm m}) mapping performance comparison.}\label{MOSPA}
	\begin{threeparttable}	
		\begin{tabular}{cccc}
			\toprule
			\multirow{2}{*}{Bias Level}& $\omega_n^{(m)}=1$ m  & $\omega_n^{(m)}=5$ m & $\omega_n^{(m)}=10$ m \\
			& $\alpha_n=0.1$ rad & $\alpha_n=0.2$ rad & $\alpha_n=0.5$ rad \\
			\hline
			Proposed Method & {\bf 1.38} & {\bf 1.85} & {\bf 2.41} \\
			Without Bias Est. & 2.86 & 6.57 & 9.92 \\	
			\bottomrule			
		\end{tabular}
	\end{threeparttable}	
\end{table}

	Through combining different categories of measurements, we obtain better SLAM results than those from a single measurement category, thereby verifying the performance gain of the proposed measurement plug-and-play mechanism.
	However, a single agent cannot obtain the entire features in the ROI because of the feature loss caused by measurement biases, such as the clock bias (the subfigure in the upper right corner of Fig. \ref{fig:w1}).  
	Therefore, we explore the performance of the proposed crowdsourcing mechanism for multi-agent collaboration in the following subsection.

	\subsection{Performance of the Probabilistic Crowdsourcing Mechanism}

	\subsubsection{Case 1 - Single category of measurements is used with agents from different entrances}

\begin{figure}
		\vspace{-0.2cm}
		\centering
		\includegraphics[scale=0.5]{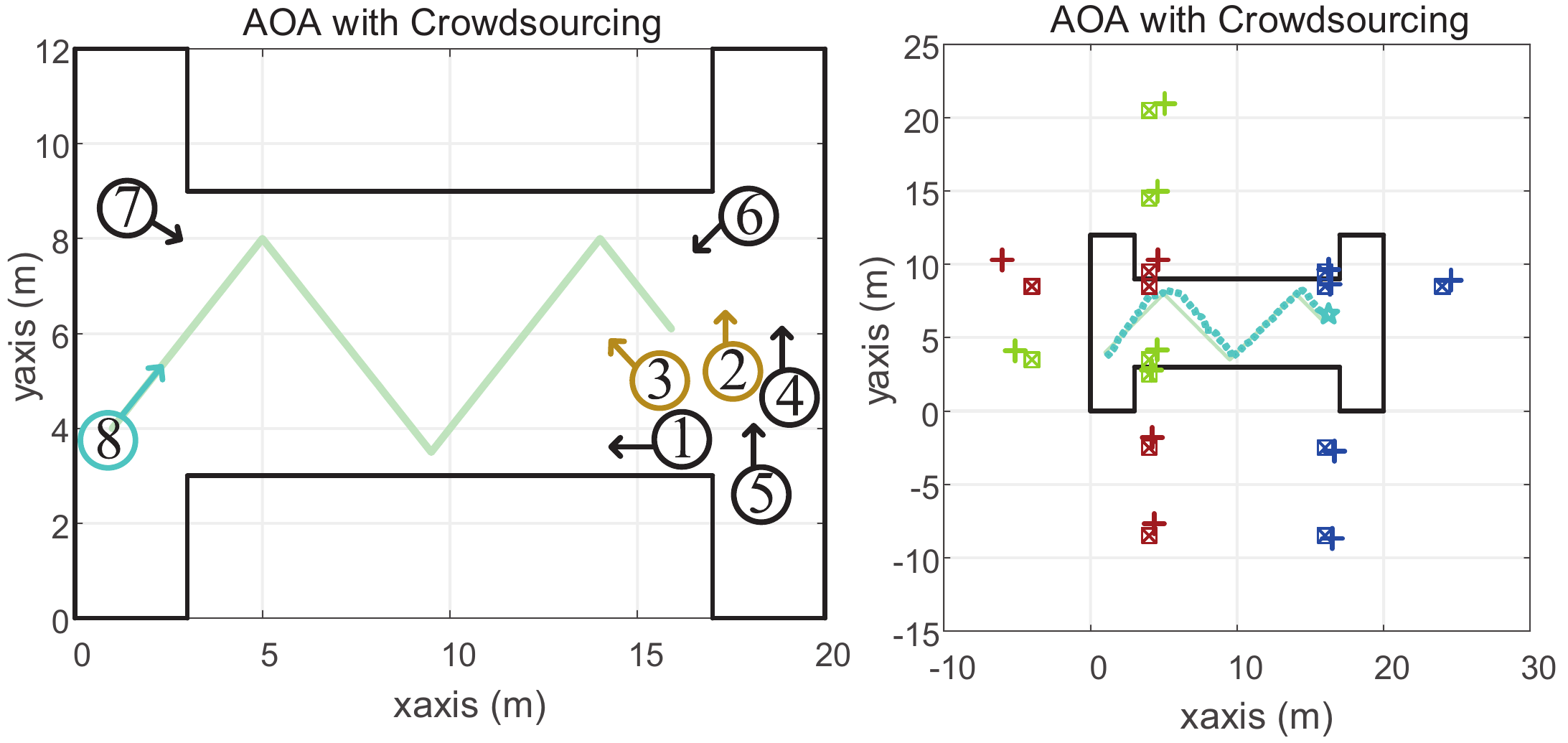}
		\caption{SLAM results for Case 1 with the proposed crowdsourcing mechanism.}\label{fig:cs-a}
	\end{figure}

	We consider eight agents, and the start point of each agent is shown in the left subfigure of Fig. \ref{fig:cs-a}. 
	We divide the eight agents into three batches.
	Agents $1$, $4$, $5$, $6$, and $7$ constitute the first batch, which enters the ROI at time slot $n=1$.
	Agents $2$ and $3$ are in the second batch, which enters the ROI at time slot $n=5$.
	Agent $8$ enters the ROI at time slot $n=10$.
	Only the AOA measurements are used by each agent.
	The SLAM result of Agent $8$ is shown in the right subfigure of Fig. \ref{fig:cs-a}, where the average positioning error is less than $0.5$ m and the average mapping error is less than $1$ m. With the proposed crowdsourcing mechanism in the case of agents from different entrances, the scaling phenomenon in the SLAM result caused by the AOA measurement is eliminated.

	\begin{figure}
		\vspace{-0.2cm}
		\centering
		\includegraphics[scale=0.33]{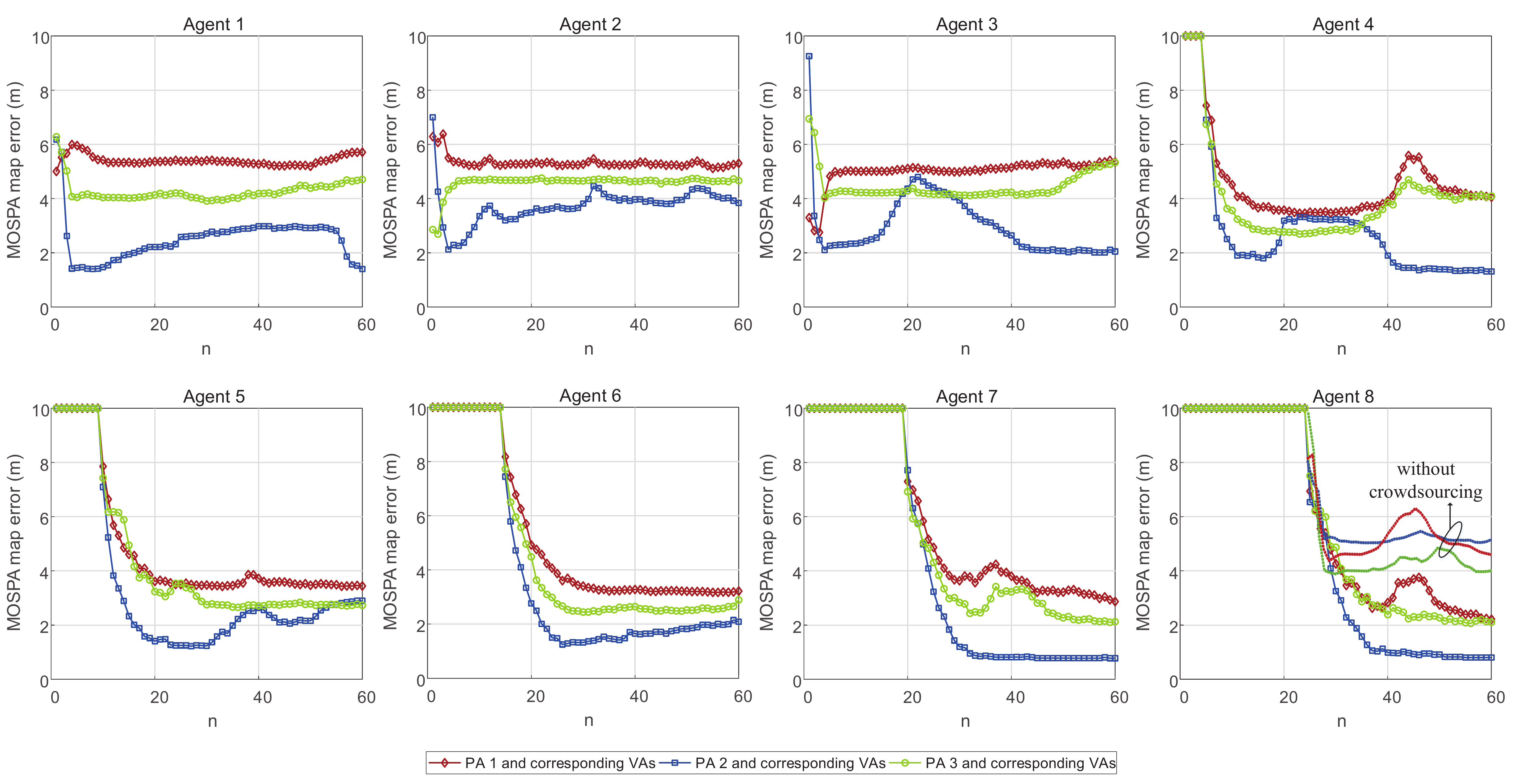}
		\caption{Mapping results obtained from
			$100$ independent Monte Carlo simulations for Case 2 with the proposed crowdsourcing mechanism.}\label{fig:cs-b} 
	\end{figure}

	\begin{figure}
	\centering
	\includegraphics[scale=0.6]{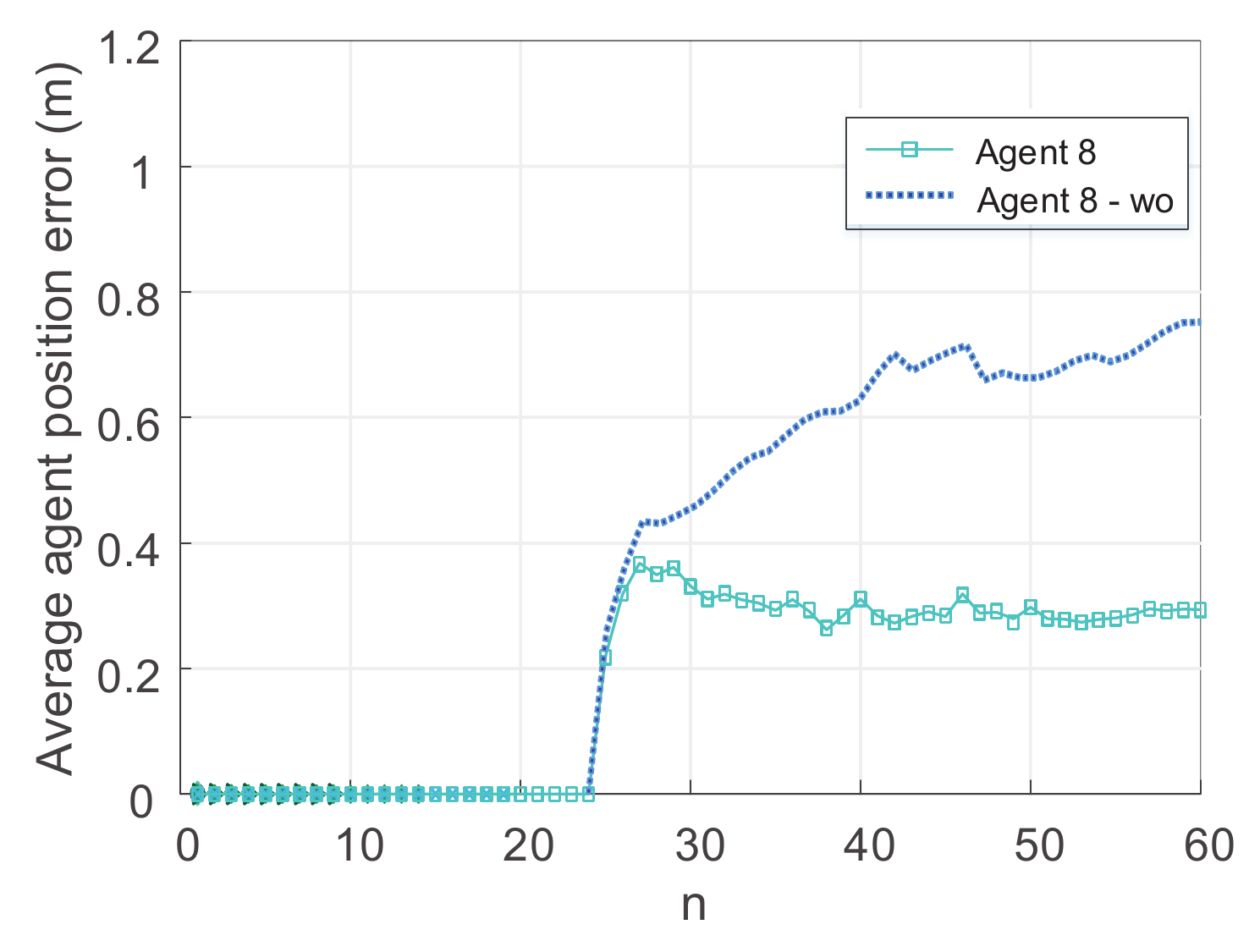}
	\caption{Positioning results obtained from
		$100$ independent Monte Carlo simulations for Case 2 with the proposed crowdsourcing mechanism.}\label{fig:cs-c} 
    \end{figure}

	\subsubsection{Case 2 - Multiple categories of measurements are used with agents from the same entrance} 
	We consider eight agents with 
	the combination of AOA and TOA measurements.
	The time of the agents enter the ROI are $n=1$ for Agents $1$, $2$, and $3$; $n=5$ for Agent $4$, $n=10$ for Agent $5$, $n=15$ for Agent $6$, $n=20$ for Agent $7$, and $n=25$ for Agent $8$.
	The total times of Monte-Carlo simulations are set to $100$.
	Fig. \ref{fig:cs-b} shows the average mapping performance
	of the proposed crowdsourcing mechanism.
	As Agents $1$, $2$, and $3$ have no prior information to inherit, the mapping error converges at approximately $4$ m.
	Then, the mapping results are gradually improved from Agents $4$ to $8$. 
	The dotted lines of Agent $8$ show the mapping result without crowdsourcing, which is approximately three times worse than those of the solid lines on average.
	Fig. \ref{fig:cs-c} shows the positioning performance of the proposed crowdsourcing mechanism.
	The average positioning accuracy of Agent $8$ improved by $42.5 \%$ at $n=60$ with the proposed crowdsourcing method.
	The simulation result verifies the effectiveness of the proposed crowdsourcing method.

	\section{Conclusion}

	This study proposed a SLAM method with measurement plug-and-play and crowdsourcing mechanisms in a Bayesian framework. First, we extended the classic BP-based SLAM method by considering practical requirements and thus realized a measurement plug-and-play function. In particular, we divided the measurements into three categories according to their unknown biases. We explained the mechanism in detail by taking TOA, AOA, and RSS measurements as examples. The simulation results showed that the proposed algorithm estimates the time-varying agent's and features' states and corresponding measurement biases (such as clock bias, orientation bias, and unknown RSS model parameters) with high accuracy and robustness in challenging scenarios. The proposed plug-and-play SLAM can do the estimates without prior information on the floor plan, anchors, or agents. We also established a probabilistic crowdsourcing mechanism and designed a corresponding physical layer frame structure. Our study is the first BP-based crowdsourcing solving ``double count" and ``data reliability" problems through probabilistic data association's flexible application. Our mechanism does not require the location information of agents for privacy and security considerations. Simulation results revealed that the proposed crowdsourcing mechanism could further enhance the plug-and-play SLAM's performance through cooperation among agents.

	\begin{appendices}
		
		\section{}\label{A}
		In this subsection, we derive the factorization of data association pdf in \eqref{da1}.  First, according to \cite{BP1}, we have
		\begin{equation}\label{da0}
		\begin{array}{ll}
		&f(\mathbf{a}_{n}^{(m)},\mathbf{b}_{n}^{(m)},{c}_{n}^{(m)},\breve{{\mathbf{v}}}_{n}^{(m)}\big|\tilde{{\mathbf{v}}}_{n}^{(m)},\mathbf{u}_{n},{\alpha}_{n},\bm{\omega}_{n}^{(m)})  \\
		= & p(\mathbf{a}_{n}^{(m)},\mathbf{b}_{n}^{(m)},{c}_{n}^{(m)},\breve{{\mathbf{r}}}_{n}^{(m)}\big|\tilde{{\mathbf{v}}}_{n}^{(m)},\mathbf{u}_{n},{\alpha}_{n},\bm{\omega}_{n}^{(m)})  f(\breve{\mathbf{{v}}}_{n}^{(m)}\big|\breve{\mathbf{{r}}}_{n}^{(m)},\mathbf{u}_{n},{\alpha}_{n},\bm{\omega}_{n}^{(m)},{c}_{n}^{(m)}),
		\end{array} 
		\end{equation}
		where $\breve{{\mathbf{r}}}_{n}^{(m)}=[\breve{{\mathbf{r}}}_{n,1}^{(m)},\ldots,\breve{{\mathbf{r}}}_{n,|\mathcal{N}_{n}^{(m)}|}^{(m)}]$.
		Since $\mathbf{a}_{n}^{(m)}$ determines $\mathbf{b}_{n}^{(m)}$, we have
		\begin{equation}\label{mmp5}
		p(\mathbf{a}_{n}^{(m)},\mathbf{b}_{n}^{(m)},{c}_{n}^{(m)},\breve{{\mathbf{r}}}_{n}^{(m)}\big|\tilde{{\mathbf{v}}}_{n}^{(m)},\mathbf{u}_{n},{\alpha}_{n},\bm{\omega}_{n}^{(m)})    
		\propto	 p(\mathbf{a}_{n}^{(m)},{c}_{n}^{(m)},\breve{{\mathbf{r}}}_{n}^{(m)}\big|\tilde{{\mathbf{v}}}_{n}^{(m)},\mathbf{u}_{n},{\alpha}_{n},\bm{\omega}_{n}^{(m)}). 
		\end{equation}
		Let $\bm{\mu}_n^{(m)}$ denote the indicate vector of $\mathbf{a}_{n}^{(m)}$, where ${\mu}_{n,i}^{(m)}=1$ when ${a}_{n,i}^{(m)}\neq 0$; and ${\mu}_{n,i}^{(m)}=0$ when ${a}_{n,i}^{(m)} = 0$. 
		Therefore, we have 
		\begin{equation}\label{iindi3}
		\hspace{-0.35cm}p(\mathbf{a}_{n}^{(m)},{c}_{n}^{(m)},\breve{{\mathbf{r}}}_{n}^{(m)}\big|\tilde{{\mathbf{v}}}_{n}^{(m)},\mathbf{u}_{n},{\alpha}_{n},\bm{\omega}_{n}^{(m)})
		\!\!=\!\! p(\mathbf{a}_{n}^{(m)},\breve{\mathbf{{r}}}_{n}^{(m)},\bm{\mu}_n^{(m)},|\mathcal{N}_{n}^{(m)}|,|\mathcal{F}_{n}^{(m)}|\big|\tilde{{\mathbf{v}}}_{n}^{(m)},\mathbf{u}_{n},{\alpha}_{n},\bm{\omega}_{n}^{(m)}).
		\end{equation}
		According to the chain rule, we obtain
		\begin{equation}\label{appen1}
		\begin{array}{ll}
		&\!\!\!\!\!p(\mathbf{a}_{n}^{(m)},\breve{\mathbf{{r}}}_{n}^{(m)},\bm{\mu}_n^{(m)},|\mathcal{N}_{n}^{(m)}|,|\mathcal{F}_{n}^{(m)}|\big|\tilde{{\mathbf{v}}}_{n}^{(m)},\mathbf{u}_{n},{\alpha}_{n},\bm{\omega}_{n}^{(m)})\\
		\!\!\!=&\!\!\!\!\!\Psi(\mathbf{a}_{n}^{(m)},\mathbf{b}_{n}^{(m)}) p(\breve{\mathbf{{r}}}_{n}^{(m)}\big|\mathbf{a}_{n}^{(m)},\bm{\mu}_n^{(m)},|\mathcal{N}_{n}^{(m)}|,|\mathcal{F}_{n}^{(m)}|,\tilde{{\mathbf{v}}}_{n}^{(m)},\mathbf{u}_{n},{\alpha}_{n},\bm{\omega}_{n}^{(m)})\\
		&\!\!\!\!\!\times p(\mathbf{a}_{n}^{(m)}\big|\bm{\mu}_n^{(m)},|\mathcal{N}_{n}^{(m)}|,|\mathcal{F}_{n}^{(m)}|,\tilde{{\mathbf{v}}}_{n}^{(m)},\mathbf{u}_{n},{\alpha}_{n},\bm{\omega}_{n}^{(m)}) p(\bm{\mu}_n^{(m)}\big|\tilde{{\mathbf{v}}}_{n}^{(m)},\mathbf{u}_{n},{\alpha}_{n},\bm{\omega}_{n}^{(m)})p(|\mathcal{N}_{n}^{(m)}|)p(|\mathcal{F}_{n}^{(m)}|).\\
		\end{array}
		\end{equation}
		As $\breve{{\mathbf{r}}}_{n}^{(m)}$ is a $|\mathcal{N}_{n}^{(m)}|$-dimensional vector, given $\bm{\mu}_n^{(m)}$, $|\mathcal{N}_{n}^{(m)}|$, and $|\mathcal{F}_{n}^{(m)}|$ ($|\mathcal{D}_{n}^{(m)}|$ and $|\mathcal{M}_{n}^{(m)}|$ can be known), there are ${{\rm C }^{ |\mathcal{N}_{n}^{(m)}|}_{|\mathcal{M}_{n}^{(m)}|-|\mathcal{D}_{n}^{(m)}|}}$ kinds of combination of $\breve{{\mathbf{r}}}_{n}^{(m)}$.
		Hence, the probability is given as
		\begin{equation}\label{appen2}
		\begin{array}{ll}
		\!\!\!\!\!\!&\!\!\!p(\breve{\mathbf{{r}}}_{n}^{(m)}\big|\mathbf{a}_{n}^{(m)},\bm{\mu}_n^{(m)},|\mathcal{N}_{n}^{(m)}|,|\mathcal{F}_{n}^{(m)}|,\tilde{{\mathbf{v}}}_{n}^{(m)},\mathbf{u}_{n},{\alpha}_{n},\bm{\omega}_{n}^{(m)})= \dfrac{|\mathcal{N}_{n}^{(m)}|!|\mathcal{F}_{n}^{(m)}| !}{(|\mathcal{M}_{n}^{(m)}|-|\mathcal{D}_{n}^{(m)}| )!}.
		\end{array}
		\end{equation}
		Note that $\mathbf{a}_{n}^{(m)}$ is an ordered vector.
		Given $\bm{\mu}_n^{(m)}$, $|\mathcal{N}_{n}^{(m)}|$, and $|\mathcal{F}_{n}^{(m)}|$ ($|\mathcal{D}_{n}^{(m)}|$ and $|\mathcal{M}_{n}^{(m)}|$ can be known),
		there are ${{\rm A }^{|\mathcal{D}_{n}^{(m)}|}_{|\mathcal{M}_{n}^{(m)}|}}$ kinds of combination of $\mathbf{a}_{n}^{(m)}$, then, we obtain \vspace{-0.2cm}
		\begin{equation}\label{indi6}
		\begin{array}{ll}
		\!\!\!\!\!\!\!\!\!\!&\!\!\!p(\mathbf{a}_{n}^{(m)}\big|\bm{\mu}_n^{(m)},|\mathcal{N}_{n}^{(m)}|,|\mathcal{F}_{n}^{(m)}|,\tilde{{\mathbf{v}}}_{n}^{(m)},\mathbf{u}_{n},{\alpha}_{n},\bm{\omega}_{n}^{(m)})=\dfrac{(|\mathcal{M}_{n}^{(m)}|-|\mathcal{D}_{n}^{(m)}| )!}{|\mathcal{M}_{n}^{(m)}|!}.\vspace{-0.2cm}
		\end{array}
		\end{equation}
		As $\bm{\mu}_n^{(m)}$ is an indicator vector of the legacy features, it is determined by the detection probability of a feature as\vspace{-0.2cm}
		\begin{equation}\label{indi7} 
		\hspace{-0.2cm}
		\begin{array}{ll}
		&\!\!\!\!p(\bm{\mu}_n^{(m)}\big|\tilde{{\mathbf{v}}}_{n}^{(m)},\mathbf{u}_{n},{\alpha}_{n},\bm{\omega}_{n}^{(m)})=\!\! \prod\limits_{j\in\mathcal{D}_{n}^{(m)}}P_d^{(m)}(\mathbf{u}_{n},\mathbf{{p}}_{n,{a}_{n,j}^{(m)}}^{(m)}) \prod\limits_{j'\in\bar{\mathcal{D}}_{n}^{(m)}}\left(1-P_d^{(m)}(\mathbf{u}_{n},\mathbf{{p}}_{n,j'}^{(m)})\right).\vspace{-0.2cm}
		\end{array}
		\end{equation}
		The number of newly detected features and false alarms follows Poisson distributions, therefore the probabilities are given as \vspace{-0.2cm}
		\begin{equation}\label{indi8} 
		p(|\mathcal{N}_{n}^{(m)}|) =  {e^{-\mu_{\rm new}^{(m)}}\mu_{\rm new}^{(m)|\mathcal{N}_{n}^{(m)}|}}/{|\mathcal{N}_{n}^{(m)}|!},\vspace{-0.2cm}
		\end{equation}
		and\vspace{-0.2cm}
		\begin{equation}\label{indi9} 
		p(|\mathcal{F}_{n}^{(m)}|) = {e^{-\mu_{\rm false}^{(m)}}\mu_{\rm false}^{(m)|\mathcal{F}_{n}^{(m)}|}}/{|\mathcal{F}_{n}^{(m)}|!},\vspace{-0.2cm}
		\end{equation}
		respectively.
		The prior pdf of the state of new features conditioned on the existence indicator of new features, the state of the agent, and the number of measurements is given by 	
		\begin{equation}\label{mmp3}
		\begin{array}{ll}
		&f(\breve{\mathbf{{v}}}_{n}^{(m)}|\breve{\mathbf{{r}}}_{n}^{(m)},\mathbf{u}_{n},{\alpha}_{n},\bm{\omega}_{n}^{(m)},{c}_{n}^{(m)}) = \prod\limits_{k\in\mathcal{N}_{n}^{(m)}}  f_{\rm new}(\breve{{\mathbf{v}}}_{n,k}^{(m)}|\mathbf{u}_{n},{\alpha}_{n},\bm{\omega}_{n}^{(m)})\prod\limits_{k'\in\bar{\mathcal{N}}_{n}^{(m)}}  f_D(\breve{{\mathbf{v}}}_{n,k'}^{(m)}).
		\end{array}
		\end{equation}
		Finally, 
		by multiplying \eqref{appen2}-\eqref{mmp3} and $\Psi(\mathbf{a}_{n}^{(m)},\mathbf{b}_{n}^{(m)})$, we can obtain \eqref{da1}.

	\end{appendices}
	
	\bibliographystyle{IEEEtran}
	\bibliography{bibsample}

\begin{thebibliography}{10}
\providecommand{\url}[1]{#1}
\csname url@samestyle\endcsname
\providecommand{\newblock}{\relax}
\providecommand{\bibinfo}[2]{#2}
\providecommand{\BIBentrySTDinterwordspacing}{\spaceskip=0pt\relax}
\providecommand{\BIBentryALTinterwordstretchfactor}{4}
\providecommand{\BIBentryALTinterwordspacing}{\spaceskip=\fontdimen2\font plus
\BIBentryALTinterwordstretchfactor\fontdimen3\font minus
  \fontdimen4\font\relax}
\providecommand{\BIBforeignlanguage}[2]{{%
\expandafter\ifx\csname l@#1\endcsname\relax
\typeout{** WARNING: IEEEtran.bst: No hyphenation pattern has been}%
\typeout{** loaded for the language `#1'. Using the pattern for}%
\typeout{** the default language instead.}%
\else
\language=\csname l@#1\endcsname
\fi
#2}}
\providecommand{\BIBdecl}{\relax}
\BIBdecl

\bibitem{6G1}
\BIBentryALTinterwordspacing
M.~{Latva-aho} and K.~{Lepp\"{a}nen}, ``Key drivers and research challenges for
  6{G} ubiquitous wireless intelligence,'' 2019. [Online]. Available:
  \url{http://urn.fi/urn:isbn:9789526223544}
\BIBentrySTDinterwordspacing

\bibitem{BP2}
E.~{Leitinger}, F.~{Meyer}, F.~{Hlawatsch}, K.~{Witrisal}, F.~{Tufvesson}, and
  M.~Z. {Win}, ``A belief propagation algorithm for multipath-based {SLAM},''
  \emph{IEEE Trans. Wireless Commun.}, vol.~18, no.~12, pp. 5613--5629, Sept.
  2019.

\bibitem{Location-aware}
R.~{Di Taranto}, S.~{Muppirisetty}, R.~{Raulefs}, D.~{Slock}, T.~{Svensson},
  and H.~{Wymeersch}, ``Location-aware communications for 5{G} networks: How
  location information can improve scalability, latency, and robustness of
  5{G},'' \emph{IEEE Signal Process. Mag.}, vol.~31, no.~6, pp. 102--112, Oct.
  2014.

\bibitem{UWB}
E.~{Leitinger}, P.~{Meissner}, C.~{Rüdisser}, G.~{Dumphart}, and
  K.~{Witrisal}, ``Evaluation of position-related information in multipath
  components for indoor positioning,'' \emph{IEEE J. Sel. Areas Commun.},
  vol.~33, no.~11, pp. 2313--2328, May 2015.

\bibitem{SOO}
C.~{Gentner}, T.~{Jost}, W.~{Wang}, S.~{Zhang}, A.~{Dammann}, and U.~{Fiebig},
  ``Multipath assisted positioning with simultaneous localization and
  mapping,'' \emph{IEEE Trans. Wireless Commun.}, vol.~15, no.~9, pp.
  6104--6117, Jun. 2016.

\bibitem{RSS}
J.~{Choi} and Y.~{Choi}, ``Calibration-free positioning technique using
  {W}i-{F}i ranging and built-in sensors of mobile devices,'' \emph{IEEE
  Internet Things J.}, pp. 1--1, Jun. 2020.

\bibitem{beaconCSI}
\BIBentryALTinterwordspacing
J.~{Choi}, ``Sensor-aided learning for {W}i-{F}i positioning with beacon
  channel state information,'' 2020, in press. [Online]. Available:
  \url{https://arxiv.org/abs/2007.06204}
\BIBentrySTDinterwordspacing

\bibitem{loc1}
F.~{Lemic}, J.~{Martin}, C.~{Yarp}, D.~{Chan}, V.~{Handziski}, R.~{Brodersen},
  G.~{Fettweis}, A.~{Wolisz}, and J.~{Wawrzynek}, ``Localization as a feature
  of mm{W}ave communication,'' in \emph{proc. IWCMC}, Sept. 2016, pp.
  1033--1038.

\bibitem{mmwave1}
H.~{Wymeersch}, G.~{Seco-Granados}, G.~{Destino}, D.~{Dardari}, and
  F.~{Tufvesson}, ``5{G} mm{W}ave positioning for vehicular networks,''
  \emph{IEEE Wireless Commun.}, vol.~24, no.~6, pp. 80--86, Dec. 2017.

\bibitem{mmwave2}
A.~{Shahmansoori}, G.~E. {Garcia}, G.~{Destino}, G.~{Seco-Granados}, and
  H.~{Wymeersch}, ``Position and orientation estimation through millimeter-wave
  {MIMO} in 5{G} systems,'' \emph{IEEE Trans. Wireless Commun.}, vol.~17,
  no.~3, pp. 1822--1835, Dec. 2017.

\bibitem{loc2}
B.~{Zhou}, A.~{Liu}, and V.~{Lau}, ``Successive localization and beamforming in
  5{G} mm{W}ave {MIMO} communication systems,'' \emph{IEEE Trans. Signal
  Process.}, vol.~67, no.~6, pp. 1620--1635, Jan. 2019.

\bibitem{slam2}
R.~{Mendrzik}, F.~{Meyer}, G.~{Bauch}, and M.~Z. {Win}, ``Enabling situational
  awareness in millimeter wave massive {MIMO} systems,'' \emph{IEEE J. Sel.
  Topics Signal Process.}, vol.~13, no.~5, pp. 1196--1211, Aug. 2019.

\bibitem{mmwave}
T.~S. {Rappaport}, F.~{Gutierrez}, E.~{Ben-Dor}, J.~N. {Murdock}, Y.~{Qiao},
  and J.~I. {Tamir}, ``Broadband millimeter-wave propagation measurements and
  models using adaptive-beam antennas for outdoor urban cellular
  communications,'' \emph{IEEE Trans. Antennas and Propag.}, vol.~61, no.~4,
  pp. 1850--1859, Dec. 2012.

\bibitem{slam-survey}
G.~{Bresson}, Z.~{Alsayed}, L.~{Yu}, and S.~{Glaser}, ``Simultaneous
  localization and mapping: A survey of current trends in autonomous driving,''
  \emph{IEEE Trans. Intell. Veh.}, vol.~2, no.~3, pp. 194--220, Sept. 2017.

\bibitem{lc1}
B.~{Williams}, M.~{Cummins}, J.~{Neira}, P.~{Newman}, I.~{Reid}, and
  J.~{Tard\'{o}s}, ``A comparison of loop closing techniques in monocular
  {SLAM},'' \emph{Robot. Auton. Syst.}, vol.~57, no.~12, pp. 1188--1197, Feb.
  2009.

\bibitem{lc}
T.~{Naseer}, M.~{Ruhnke}, C.~{Stachniss}, L.~{Spinello}, and W.~{Burgard},
  ``Robust visual {SLAM} across seasons,'' in \emph{proc. IEEE/RSJ IROS}, Dec.
  2015, pp. 2529--2535.

\bibitem{8920098}
R.~{Liu}, S.~H. {Marakkalage}, M.~{Padmal}, T.~{Shaganan}, C.~{Yuen}, Y.~L.
  {Guan}, and U.~{Tan}, ``Collaborative {SLAM} based on {W}i{F}i fingerprint
  similarity and motion information,'' \emph{IEEE Internet Things J.}, vol.~7,
  no.~3, pp. 1826--1840, Mar. 2020.

\bibitem{llm}
C.~{Guo}, K.~{Kidono}, J.~{Meguro}, Y.~{Kojima}, M.~{Ogawa}, and T.~{Naito},
  ``A low-cost solution for automatic lane-level map generation using
  conventional in-car sensors,'' \emph{IEEE Trans. Intell. Transport. Syst.},
  vol.~17, no.~8, pp. 2355--2366, Feb. 2016.

\bibitem{CSLAM}
J.~{Palacios}, G.~{Bielsa}, P.~{Casari}, and J.~{Widmer},
  ``Communication-driven localization and mapping for millimeter wave
  networks,'' in \emph{proc. IEEE INFOCOM}, Oct. 2018, pp. 2402--2410.

\bibitem{CSLAM2}
H.~{Kim}, K.~{Granström}, L.~{Gao}, G.~{Battistelli}, S.~{Kim}, and
  H.~{Wymeersch}, ``5{G} mmwave cooperative positioning and mapping using
  multi-model {PHD} filter and map fusion,'' \emph{IEEE Trans. Wireless
  Commun.}, vol.~19, no.~6, pp. 3782--3795, Mar. 2020.

\bibitem{CSLAM3}
H.~{Wymeersch} and G.~{Seco-Granados}, ``Adaptive detection probability for
  mmwave 5{G} {SLAM},'' in \emph{proc. 6G SUMMIT}, May 2020, pp. 1--5.

\bibitem{CSLAM4}
H.~{Kim}, K.~{Granström}, L.~{Gao}, G.~{Battistelli}, S.~{Kim}, and
  H.~{Wymeersch}, ``Joint {CKF}-{PHD} filter and map fusion for 5{G} multi-cell
  {SLAM},'' in \emph{proc. IEEE ICC}, Jul. 2020, pp. 1--6.

\bibitem{BP-SLAM}
E.~{Leitinger}, S.~{Grebien}, and K.~{Witrisal}, ``Multipath-based {SLAM}
  exploiting {AOA} and amplitude information,'' in \emph{proc. IEEE ICC
  Workshops}, Jul. 2019, pp. 1--7.

\bibitem{mslam}
X.~{Chu}, Z.~{Lu}, D.~{Gesbert}, L.~{Wang}, and X.~{Wen}, ``Vehicle
  localization via cooperative channel mapping,'' \emph{IEEE Trans. Veh.
  Technol.}, vol.~70, no.~6, pp. 5719--5733, Apr. 2021.

\bibitem{crowdsourcing}
Y.~{Li}, Z.~{He}, Z.~{Gao}, Y.~{Zhuang}, C.~{Shi}, and N.~{El-Sheimy}, ``Toward
  robust crowdsourcing-based localization: A fingerprinting accuracy indicator
  enhanced wireless/magnetic/inertial integration approach,'' \emph{IEEE
  Internet Things J.}, vol.~6, no.~2, pp. 3585--3600, Dec. 2018.

\bibitem{crowdsourcing-wifi}
Y.~{Zhuang}, Z.~{Syed}, Y.~{Li}, and N.~{El-Sheimy}, ``Evaluation of two
  {W}i{F}i positioning systems based on autonomous crowdsourcing of handheld
  devices for indoor navigation,'' \emph{IEEE Trans. Mobile Computing},
  vol.~15, no.~8, pp. 1982--1995, Sept. 2015.

\bibitem{crowdsourcing-quality}
P.~{Zhang}, R.~{Chen}, Y.~{Li}, X.~{Niu}, L.~{Wang}, M.~{Li}, and Y.~{Pan}, ``A
  localization database establishment method based on crowdsourcing inertial
  sensor data and quality assessment criteria,'' \emph{IEEE Internet Things
  J.}, vol.~5, no.~6, pp. 4764--4777, Mar. 2018.

\bibitem{hy}
Y.~{Han}, T.~{Hsu}, C.~{Wen}, K.~{Wong}, and S.~{Jin}, ``Efficient downlink
  channel reconstruction for {FDD} multi-antenna systems,'' \emph{IEEE Trans.
  Wireless Commun.}, vol.~18, no.~6, pp. 3161--3176, Apr. 2019.

\bibitem{nomp}
B.~{Mamandipoor}, D.~{Ramasamy}, and U.~{Madhow}, ``Newtonized orthogonal
  matching pursuit: Frequency estimation over the continuum,'' \emph{IEEE
  Trans. Signal Process.}, vol.~64, no.~19, pp. 5066--5081, Jun. 2016.

\bibitem{yj1}
J.~{Yang}, C.~{Wen}, S.~{Jin}, and F.~{Gao}, ``Beamspace channel estimation in
  mm{W}ave systems via cosparse image reconstruction technique,'' \emph{IEEE
  Trans. Commun.}, vol.~66, no.~10, pp. 4767--4782, Oct. 2018.

\bibitem{yj2}
J.~{Yang}, S.~{Jin}, C.~{Wen}, X.~{Yang}, and M.~{Matthaiou}, ``Fast beam
  training architecture for hybrid mm{W}ave transceivers,'' \emph{IEEE Trans.
  Veh. Technol.}, vol.~69, no.~3, pp. 2700--2715, Jan. 2020.

\bibitem{BP1}
F.~{Meyer}, P.~{Braca}, P.~{Willett}, and F.~{Hlawatsch}, ``A scalable
  algorithm for tracking an unknown number of targets using multiple sensors,''
  \emph{IEEE Trans. Signal Process.}, vol.~65, no.~13, pp. 3478--3493, Mar.
  2017.

\bibitem{sumpro}
F.~R. {Kschischang}, B.~J. {Frey}, and H.~. {Loeliger}, ``Factor graphs and the
  sum-product algorithm,'' \emph{IEEE Trans. Inf. Theory}, vol.~47, no.~2, pp.
  498--519, Feb. 2001.

\bibitem{sem}
Y.~{Bar-Shalom}, T.~{Kirubarajan}, and X.-R. {Li}, \emph{Estimation With
  Applications to Tracking and Navigation}.\hskip 1em plus 0.5em minus
  0.4em\relax New York, USA: Wiley, 2002.

\bibitem{M1}
J.~{Ren}, Y.~{Zhang}, K.~{Zhang}, and X.~{Shen}, ``Exploiting mobile
  crowdsourcing for pervasive cloud services: challenges and solutions,''
  \emph{IEEE Commun. Mag.}, vol.~53, no.~3, pp. 98--105, Mar. 2015.

\bibitem{M2}
T.~Li, Y.~Chen, R.~Zhang, Y.~Zhang, and T.~Hedgpeth, ``Secure crowdsourced
  indoor positioning systems,'' in \emph{IEEE INFOCOM}, 2018, pp. 1034--1042.

\end{thebibliography}
	
\end{document}